\documentclass[%
 reprint,
superscriptaddress,
 amsmath,amssymb,
 aps,
 prx,
]{revtex4-2}

\usepackage{graphicx}% Include figure files
\graphicspath{{figures/}}
\usepackage{dcolumn}% Align table columns on decimal point
\usepackage{bm}% bold math
\usepackage{color}
\usepackage{subcaption}
\usepackage{layouts}
\usepackage{siunitx}
\usepackage{xcolor}
\usepackage{tikz}
\usepackage{hyperref}
\hypersetup{
  colorlinks,
  citecolor=blue,
  linkcolor=black,
  urlcolor=blue}
\usepackage{url}
\usepackage{pgfplotstable}
\usepackage{pgfplots}
\pgfplotsset{width=10cm,compat=1.9}
\newcommand{\transpose}[1]{#1^\mathrm{T}}

\newcommand{\real}[1]{\textrm{Re}\left\{#1\right\}}

\begin{document}

\preprint{APS/123-QED}
%\title{An Systems Theory Perspective on Optimal Impulse-sensing with Nanomechanical Resoantors }% Force line breaks with \\
\title{Optimal Sensing of Momentum Kicks\\with a Feedback-Controlled Nanomechanical Resonator}% Force line breaks with \\
%\thanks{A footnote to the article title}%

\author{Kaspar Schmerling}\email{schmerling@acin.tuwien.ac.at}
    \affiliation{Automation and Control Institute (ACIN), TU Wien, Gusshausstrasse 27-29, 1040 Vienna, Austria}%
\author{Hajrudin Bešić}
    \affiliation{Institute of Sensor and Actuator Systems, TU Wien, Gusshausstrasse 27-29, 1040 Vienna, Austria}%
\author{Andreas Kugi}
    \affiliation{Automation and Control Institute (ACIN), TU Wien, Gusshausstrasse 27-29, 1040 Vienna, Austria}%
    \affiliation{Austrian Institute of Technology (AIT), Gieffinggasse 4, 1220, Vienna, Austria}%
\author{Silvan Schmid}
    \affiliation{Institute of Sensor and Actuator Systems, TU Wien, Gusshausstrasse 27-29, 1040 Vienna, Austria}%
\author{Andreas Deutschmann-Olek}\email{deutschmann@acin.tuwien.ac.at}
    \affiliation{Automation and Control Institute (ACIN), TU Wien, Gusshausstrasse 27-29, 1040 Vienna, Austria}%
    
\date{\today}% It is always \today, today,
             %  but any date may be explicitly specified

\begin{abstract}
External disturbances exciting a mechanical resonator can be exploited to gain information on the environment. Many of these interactions manifest as momentum kicks, such as the recoil of residual gas, radioactive decay, or even hypothetical interactions with dark matter. These disturbances are often rare enough that they can be resolved as singular events rather than cumulated as force noise. While high-Q resonators with low masses are particularly sensitive to such momentum kicks, they will strongly excite the resonator, leading to nonlinear effects that deteriorate the sensing performance. Hence, this paper utilizes optimal estimation methods to extract individual momentum kicks from measured stochastic trajectories of a mechanical resonator kept in the linear regime through feedback control. The developed scheme is illustrated and tested experimentally using a pre-stressed SiN trampoline resonator. Apart from enhancing a wide range of sensing scenarios mentioned above, our results indicate the feasibility of novel single-molecule mass spectrometry approaches.
%\begin{description}
%\item[Usage]
%Secondary publications and information retrieval purposes.
%\item[Structure]
%You may use the \texttt{description} environment to structure your %abstract;
%use the optional argument of the \verb+\item+ command to give the %category of each item. 
%\end{description}{Zeptogram
\end{abstract}

\maketitle

\section{\label{sec:intro} Introduction}
% Nanomechanical resonators introduction
%Low-noise 
Nanomechanical resonators are increasingly applied in various fields of research and technology, including mass spectrometry \cite{reinhardt2016ultralow, hanay2012single, hanay2015inertial, demir2021adaptive}, atomic force microscopy \cite{li2007ultra,halg2021membrane,PhysRevApplied.22.044001}, nano-magnetic resonance imaging \cite{degen2009nanoscale,grob2019magnetic}, optomechanical experiments \cite{bagci2014optical, rossi2018measurement, delic2020cooling, seis2022ground, huang2024room}, and gravitational wave detection \cite{gonzfilez1994brownian}. %A crucial aspect of the effectiveness of nanomechanical resonators is their low mechanical damping, resulting in exceptional force sensitivity. 
% State estimation and control in nMRs
Historically, the development of nanomechanical systems for sensing applications has been focused primarily on enhancing their physical capabilities through optimized mechanical design. However, advanced tools from signal processing and control theory offer various methods to extract information provided by measurements \cite{wang2023beating, wieczorek2015optimal}. One group of methods of particular interest are (optimal) state estimators, such as the Kalman-Bucy filter (KBF) and its extensions \cite{kalmBu, rtsSmo, sarkka2023bayesian}. The KBF provides an optimal solution to the state estimation problem for systems with linear dynamics disturbed by Gaussian noise processes, which is characteristic of most nanomechanical systems \cite{schmid2016fundamentals} as long as they are only weakly excited.

% Jump processes and in sensing
Many sensing applications rely on detecting jump processes in dynamical systems, where a quantity of interest undergoes sudden changes at discrete points in time, affecting the system's dynamics \cite{wang2023beating}. These processes frequently occur in nanomechanical systems \cite{Mete2024, demir2021adaptive, hanay2012single, wang_mechanical_2024}. For instance, nanoscale mass additions to the measurement apparatus cause measurable shifts in the resonance frequency of the oscillators, which can be used in mass spectrometry with single molecule sensitivity \cite{sansa2020optomechanical, stassi_large-scale_2019, ruz2020effect, sage2018single, naik2009towards, bevsic2024optimized}.

% momentum kicks
Low-effective mass oscillators are particularly sensitive to momentum changes caused by particles colliding with the resonator. For example, momentum kicks due to presumably residual gas molecules in ultra-high vacuum have been reported in \cite{magrini2021real}, which was later proposed as a way of sensing pressure in ultra-high vacuum environments exceeding current limitations \cite{barker2024collision}. Furthermore, sensing of momentum kicks has been proposed to observe radioactive decay \cite{wang_mechanical_2024} and hypothetical interactions with dark matter \cite{carney_mechanical_2021} and could possibly be used to detect ultrafine airborne particles \cite{li_online_2023} that are challenging to detect with existing methods.
Finally, it could enable novel recoil-based mass spectrometry methods for individual particles that would be a crucial tool for single-cell proteomics \cite{bennett_single-cell_2023}.
The ability to accurately extract momentum kicks provided to a nanomechanical resonator is thus crucial for many interesting sensing applications. 

% Current 
The present work proposes schemes to sense momentum kicks optimally using a feedback-controlled nanoelectromechanical system (NEMS). Feedback-controlling the NEMS resonator ensures that it remains weakly excited even under strong stochastic disturbances, which is also called feedback cooling \cite{hopkins2003feedback, kleckner2006sub}. In particular, this includes the energy transferred to the resonator due to the repeated external momentum kicks we aim to estimate.
Thus, the measured behavior of the NEMS resonator can be accurately described as a linear dynamic system driven by stochastic disturbances while observing a measurement output tainted by noise and other imperfections.
Using this mathematical description, an optimal algorithm to estimate individual momentum kicks is derived by combining Kalman-Bucy filtering and smoothing methods. The latter are directly related to so-called retrodiction methods \cite{zhang_prediction_2017,bao_retrodiction_2020,lammers_quantum_2024} that recently gained attention in the quantum physics community.
Finally, we demonstrate and evaluate the proposed method on a high\nobreakdash-Q pre-stressed silicon nitride trampoline resonator with pronounced multi-mode behavior.

%implementing an equivalent measurement procedure for nMR systems could pave the way for new applications such as impulse-based single molecule mass spectrometry or residual pressure gauges 

%Signal processing and filtering in the nMR community are still predominantly frequency-based methods, such as frequency estimation or power spectrum analysis \cite{schmid2016fundamentals}.
%However, time-based probabilistic filtering methods, such as the KBF, might offer tools to address many challenges in nMR research for filtering applications. We propose a time-based optimal methodology for filtering jump processes near the fundamental detection limit and demonstrate the method on a state-of-the-art nMR setup. Since the proposed method is a purely filtering-based procedure, it is potentially applicable to a wide range of setups and applications of existing experiments.

\section{A Systems Theory Perspective on Nanomechanical Resonators} \label{chap:ssnems}
The application of methods from model-based systems and control theory requires a \emph{sufficiently} accurate mathematical description of the system. This is particularly true for optimal methods that combine model knowledge and measurement information to achieve a desired objective in the best possible manner.
The core element of this approach is the (stochastic) temporal evolution of the system's state $\mathbf{x} \in \mathbb{R}^n$ given by
\begin{equation} \label{eqn:LTI1}
\frac{\mathrm{d}}{\mathrm{d}t}\mathbf{x} = \mathbf{A}\mathbf{x} + \mathbf{B}\mathbf{u} + \mathbf{G}\boldsymbol{\eta}\textrm{,}\\
\end{equation}
with the dynamic matrix $\mathbf{A}$, the adjustable input $\mathbf{u} \in \mathbb{R}^l$ acting on the system via the input matrix $\mathbf{B}$, and the unknown disturbance input $\boldsymbol{\eta}\in\mathbb{R}^{m}$ with the disturbance matrix $\mathbf{G}$. The initial state is given by $\mathbf{x}(0) = \mathbf{x}_0$.
Furthermore, information on the state is only accessible via the measurable quantity $\mathbf{y}\in\mathbb{R}^m$ given by
\begin{equation} \label{eqn:LTI2}
\mathbf{y} = \mathbf{C}\mathbf{x} + \boldsymbol{\nu}
\end{equation}
with the measurement matrix $\mathbf{C}$ and the measurements noise $\boldsymbol{\nu}$.
We assume that $\boldsymbol{\eta}$ and $\boldsymbol{\nu}$ are white Gaussian noise processes with
\begin{subequations}\label{eqn:noises}
	\begin{equation}\label{eqn:noiseProc}
		\mathbb{E}[\boldsymbol{\eta}(t)] = \mathbf{0}, \quad  \mathbb{E}[\boldsymbol{\eta}(t)\boldsymbol{\eta}(t')] = \mathbf{I}\delta(t-t')
	\end{equation}
\begin{equation}\label{eqn:noiseMeas}
	\mathbb{E}[\boldsymbol{\nu}(t)] = \mathbf{0}, \quad  \mathbb{E}[\boldsymbol{\nu}(t)\boldsymbol{\nu}(t')] = \mathbf{R}\delta(t-t')
\end{equation}
\end{subequations}
where $\mathbf{I}$ is the identity matrix, $\mathbf{R}$ is a positive definite matrix, and $\delta$ denotes the Dirac delta function. With these properties, it is ensured that all variables from (\ref{eqn:LTI1}) and (\ref{eqn:LTI2}) remain Gaussian \cite{badawi1979stochastic}.

\subsection{Modeling of the Nanomechanical Resonator} \label{sec:modeling}
\begin{figure*}[hbt!]
	\centering
	\begin{subfigure}[l]{0.45\textwidth}
		\caption{}
		\centering
		\def\svgwidth{1.1\textwidth}
		%% Creator: Inkscape 1.2.2 (732a01da63, 2022-12-09), www.inkscape.org
%% PDF/EPS/PS + LaTeX output extension by Johan Engelen, 2010
%% Accompanies image file 'resonator_inception.pdf' (pdf, eps, ps)
%%
%% To include the image in your LaTeX document, write
%%   \input{<filename>.pdf_tex}
%%  instead of
%%   \includegraphics{<filename>.pdf}
%% To scale the image, write
%%   \def\svgwidth{<desired width>}
%%   \input{<filename>.pdf_tex}
%%  instead of
%%   \includegraphics[width=<desired width>]{<filename>.pdf}
%%
%% Images with a different path to the parent latex file can
%% be accessed with the `import' package (which may need to be
%% installed) using
%%   \usepackage{import}
%% in the preamble, and then including the image with
%%   \import{<path to file>}{<filename>.pdf_tex}
%% Alternatively, one can specify
%%   \graphicspath{{<path to file>/}}
%% 
%% For more information, please see info/svg-inkscape on CTAN:
%%   http://tug.ctan.org/tex-archive/info/svg-inkscape
%%
\begingroup%
  \makeatletter%
  \providecommand\color[2][]{%
    \errmessage{(Inkscape) Color is used for the text in Inkscape, but the package 'color.sty' is not loaded}%
    \renewcommand\color[2][]{}%
  }%
  \providecommand\transparent[1]{%
    \errmessage{(Inkscape) Transparency is used (non-zero) for the text in Inkscape, but the package 'transparent.sty' is not loaded}%
    \renewcommand\transparent[1]{}%
  }%
  \providecommand\rotatebox[2]{#2}%
  \newcommand*\fsize{\dimexpr\f@size pt\relax}%
  \newcommand*\lineheight[1]{\fontsize{\fsize}{#1\fsize}\selectfont}%
  \ifx\svgwidth\undefined%
    \setlength{\unitlength}{480.48001099bp}%
    \ifx\svgscale\undefined%
      \relax%
    \else%
      \setlength{\unitlength}{\unitlength * \real{\svgscale}}%
    \fi%
  \else%
    \setlength{\unitlength}{\svgwidth}%
  \fi%
  \global\let\svgwidth\undefined%
  \global\let\svgscale\undefined%
  \makeatother%
  \begin{picture}(1,0.75024976)%
    \lineheight{1}%
    \setlength\tabcolsep{0pt}%
    \put(0,0){\includegraphics[width=\unitlength,page=1]{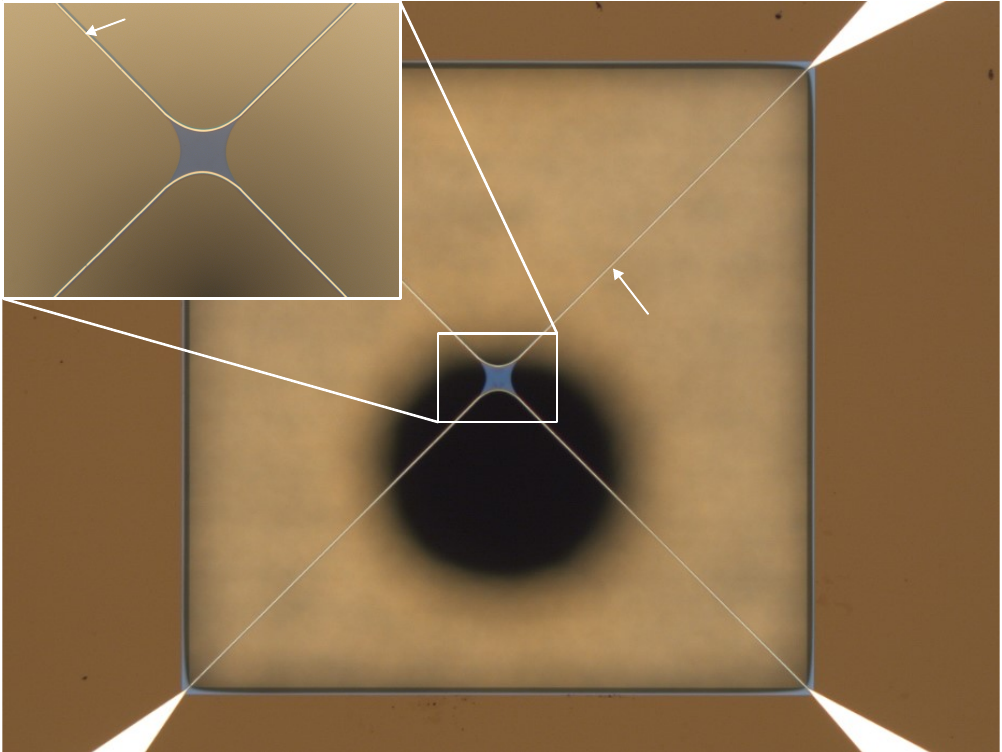}}%
    \put(0.12666053,0.71910814){\color[rgb]{1,1,1}\makebox(0,0)[lt]{\lineheight{1.25}\smash{\begin{tabular}[t]{l}Gold Wires\end{tabular}}}}%
    \put(0.65010712,0.42350932){\color[rgb]{1,1,1}\makebox(0,0)[lt]{\lineheight{1.25}\smash{\begin{tabular}[t]{l}Tethers\\\end{tabular}}}}%
    \put(0,0){\includegraphics[width=\unitlength,page=2]{resonator_inception.pdf}}%
    \put(0.87220348,0.05383206){\color[rgb]{1,1,1}\makebox(0,0)[lt]{\lineheight{1.25}\smash{\begin{tabular}[t]{l}$200\text{\textmu m}$\end{tabular}}}}%
  \end{picture}%
\endgroup%

		\label{fig:resonator_picture}
	\end{subfigure}
	\hfill
	\begin{subfigure}[l]{0.45\textwidth}
		\caption{}
		\centering
		\def\svgwidth{\textwidth}
		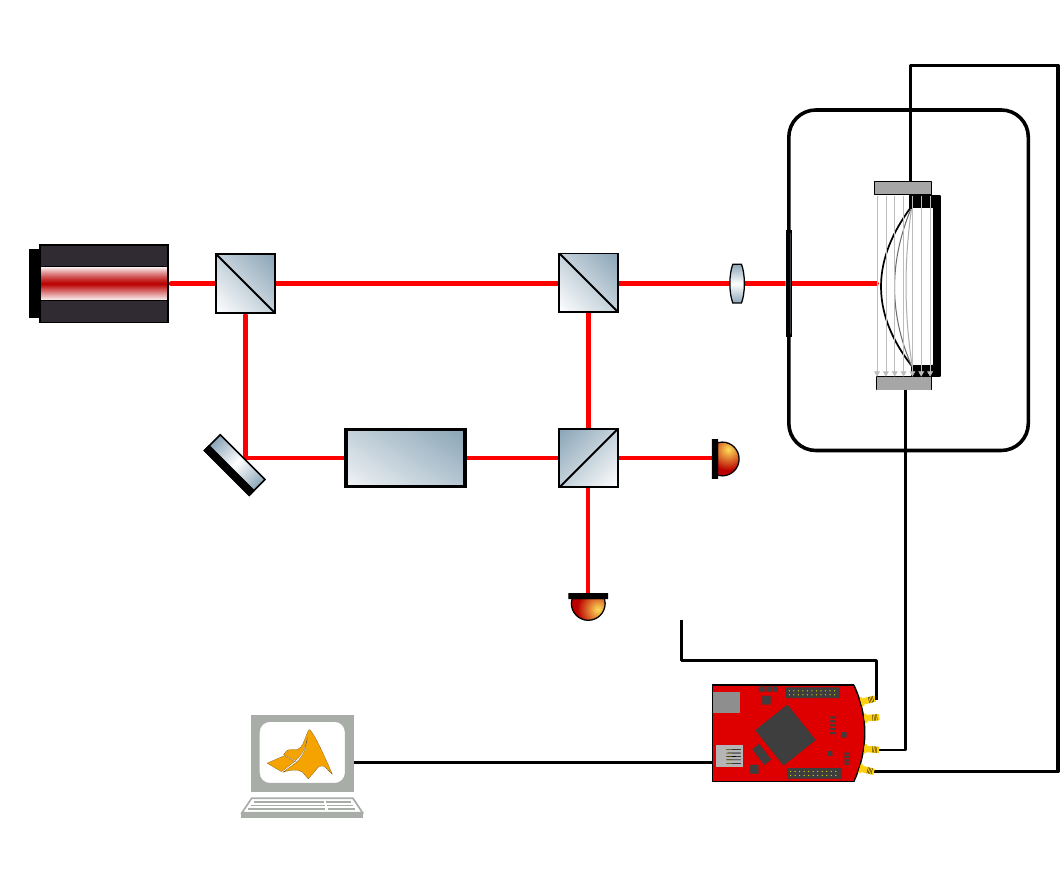
		\label{fig:expSetup1}
	\end{subfigure}
	\hfill
	\begin{subfigure}[l]{1\textwidth}
		\caption{}
		\centering
		% This file was created by matlab2tikz.
%
%The latest updates can be retrieved from
%  http://www.mathworks.com/matlabcentral/fileexchange/22022-matlab2tikz-matlab2tikz
%where you can also make suggestions and rate matlab2tikz.
%
\definecolor{mycolor1}{rgb}{0.00000,0.44700,0.74100}%
\definecolor{mycolor2}{rgb}{0.24314,0.58824,0.31765}%
\definecolor{mycolor3}{rgb}{0.85000,0.32500,0.09800}%
\begin{tikzpicture}

\begin{axis}[%
width=\textwidth,
height=0.33\textwidth,
%at={(2.708in,0.688in)},
%scale only axis,
axis on top,
xmin=1,
xmax=180,
xlabel style={font=\color{white!15!black}},
xlabel={$\omega/2\pi\quad (\SI{}{kHz})$},
ymode=log,
ymin=1e-18,
ymax=1e-08,
yminorticks=true,
ylabel style={font=\color{white!15!black}},
ylabel={$ S_{yy}(\omega)$$(\SI{}{m}^2/\SI{}{s}^2\SI{}{Hz})$},
axis background/.style={fill=white},
legend style={at={(1,0)}, anchor=south east, legend cell align=left, align=left, draw=white!15!black}
]
\addplot [color=mycolor1, forget plot]
  table[]{figures/mode_spec/mode_spec-1.tsv};
\addplot [color=mycolor1]
  table[]{figures/mode_spec/mode_spec-2.tsv};
\addlegendentry{$S_{yy}$ (measured)}

\addplot [color=mycolor2, line width=1pt]
  table[]{figures/mode_spec/mode_spec-3.tsv};
\addlegendentry{$S_{yy}$ (modeled)}

\addplot [color=mycolor3, dashed, line width=1pt]
  table[]{figures/mode_spec/mode_spec-4.tsv};
\addlegendentry{$S_{\textrm{nn}}$ (disturbance model)}

\addplot [color=white!15!black, dotted, forget plot]
  table[]{figures/mode_spec/mode_spec-5.tsv};
\addplot [color=white!15!black, dotted, forget plot]
  table[]{figures/mode_spec/mode_spec-6.tsv};
\addplot [color=white!15!black, dotted, forget plot]
  table[]{figures/mode_spec/mode_spec-7.tsv};
\addplot [forget plot] graphics [xmin=24.9896157840083, xmax=65.0103842159917, ymin=1.34590163934426e-12, ymax=8.00654098360656e-10] {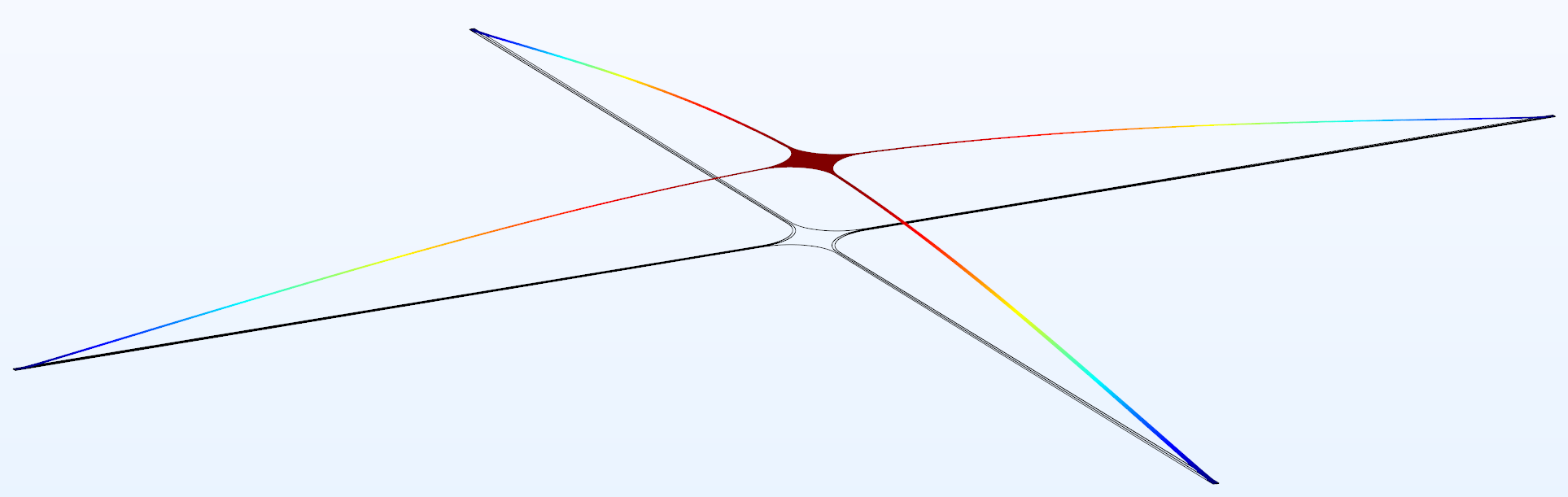};
\addplot [forget plot] graphics [xmin=69.9896373056995, xmax=110.010362694301, ymin=1.43081312410842e-12, ymax=8.00569186875892e-10] {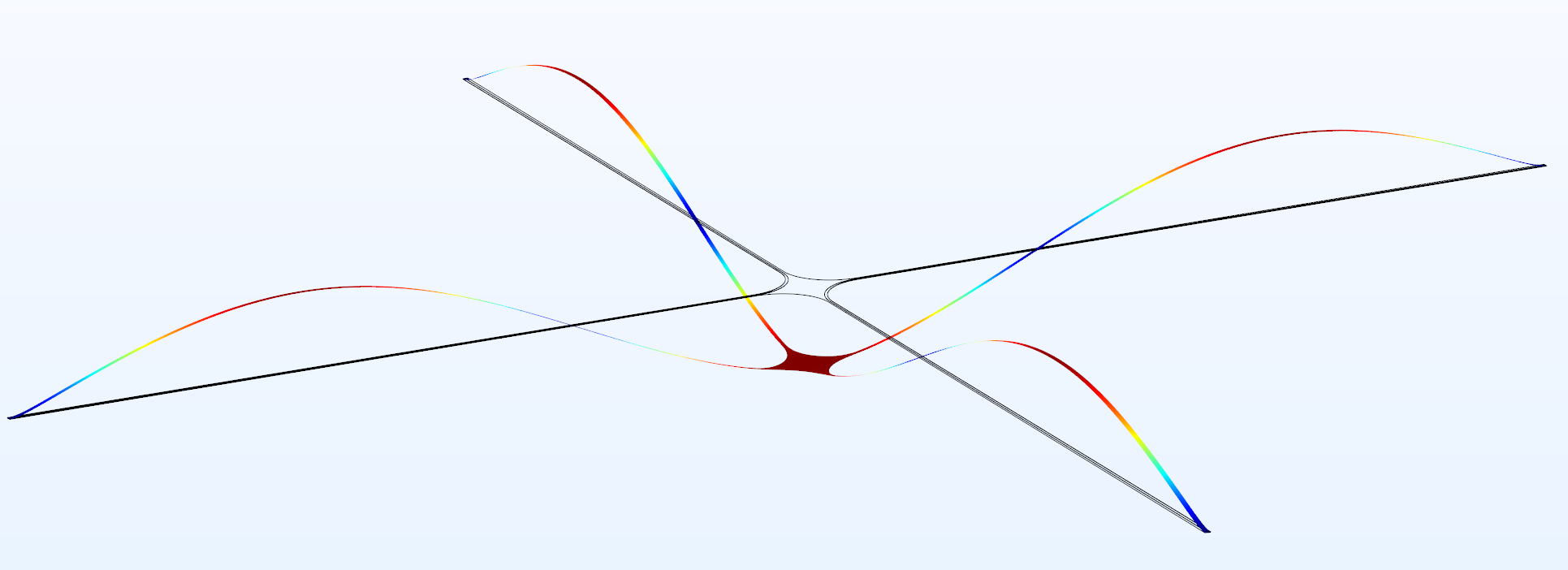};
\addplot [forget plot] graphics [xmin=114.989685404848, xmax=155.010314595152, ymin=1.43484419263456e-12, ymax=8.00565155807365e-10] {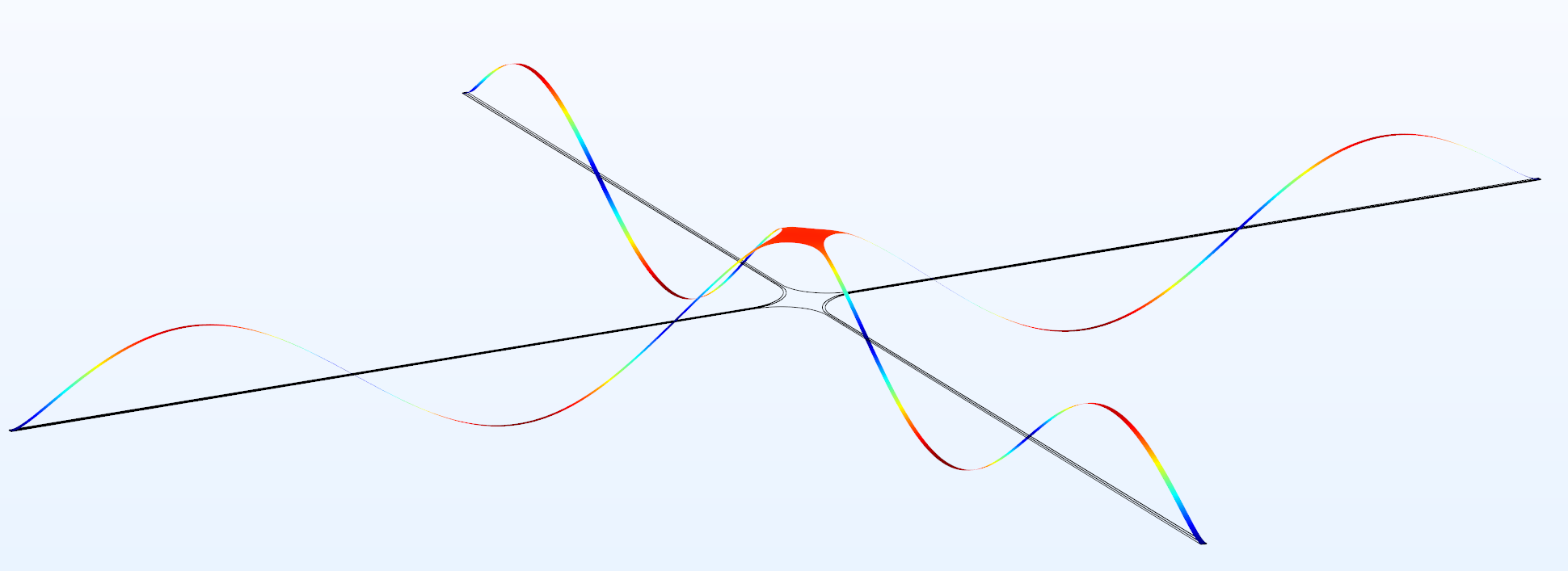};
\end{axis}

\end{tikzpicture}%
		\label{fig:ResSpec}
	\end{subfigure}
	\hfill
	\caption{(\subref{fig:resonator_picture}) Microscope picture of the used resonator. (\subref{fig:expSetup1}) Experimental setup used for interferometric velocity measurement and feedback control of the resonator: Vacuum chamber with pressure below \SI{1e-5}{mbar}. Lorentz force feedback and momentum kicks synthesized by the FPGA board. Computer for data post-processing. (\subref{fig:ResSpec}) Power spectral density of the resonator. The measured spectrum and the spectrum resulting from theory for the first three modes are depicted in blue and green, respectively. The dotted black lines represent the expected modal frequencies with their modal shapes next to them, calculated from FEM simulations.}
	\label{fig:FIG1}
\end{figure*}

In the following, a pre-stressed high-Q silicon nitride trampoline resonator is considered; see Figure~\ref{fig:FIG1}(\subref{fig:resonator_picture}) for a microscope picture. The NEMS resonator has a thickness of \SI{50}{nm}, a frame edge length of \SI{1}{mm}, a tether width of $5$\textmu\SI{}{\meter}, and a central pad measuring 45~\textmu m by 45~\textmu m. Actuation is facilitated by two gold wires, each with a thickness of \SI{200}{nm} and a width of $2.5$\textmu\SI{}{\meter}, coated onto the tethers. The resonator is immersed in a magnetic field of $\thicksim 1$T, originating from two neodymium magnets located on the sides of the resonator. Thus, the wires generate Lorenz forces acting on the resonator when a voltage signal $u(t)$ is applied. The resonator is situated within a high vacuum (HV) chamber, maintaining a residual pressure of $<$ \SI{1e-5}{mbar}. The Q-factors of the resonator modes were determined through ring-down measurements, yielding values on the order of $10^5$, with resonance frequencies in the order of tens of kilohertz.

A commercial laser-Doppler vibrometer with a laser wavelength of \SI{633}{nm}, operating in a heterodyne regime, measures the trampoline's displacement $y(t)$. The data captured by the vibrometer's photo-diodes is recorded and processed using a RedPitaya field programmable gate array (FPGA) board before being transmitted to a laboratory computer. Additionally, to apply forces to the resonator, the FPGA's outputs are connected to the resonator's transduction wires via a digital-analog converter. Figure~\ref{fig:FIG1}(\subref{fig:expSetup1}) illustrates the experimental setup. As shown in the figure, the FPGA board receives a measurement signal $y(t)$ and acts on the resonator by the feedback signal $u(t)$ and the impulse-like disturbance signal $p(t)$. 

 % Figure \ref{fig:ResSpec} shows the Power spectral density (PSD) of the measurement signal $y(t)$ in the range of \SI{0}{kHz} to \SI{180}{kHz}. The spectrum shows prominent peaks at around \SI{21}{kHz}, \SI{23}{kHz}, \SI{43}{kHz}, \SI{68}{kHz}, \SI{114}{kHz} and \SI{163}{kHz}. Also its evident, that the noise floor has a minimum at around \SI{10}{kHz} from where it increases for higher and lower frequencies. Simulating the resonator by a Finite Element Method (FEM) simulation reveals that the peaks at \SI{23}{kHz}, \SI{68}{kHz}, \SI{114}{kHz} and \SI{163}{kHz} correspond to the first four out-of-plane oscillation modes of the resonator, as the FEM simulation shows equivalent behavior. Table \ref{tab:NEMSModes} shows characteristic values of the first modes of the resonator from FEM simulation. Only modes predominantly oscillating out of plane significantly contribute to the oscillators effective mass and therefore to the overall deformation energy of the resonator. The mode shapes of the first three are depicted in Figure \ref{fig:ResSpec} next to there corresponding peaks. Additionally, the peaks at \SI{21}{kHz} and \SI{42}{kHz}, likely originate from unknown oscillation processes in the measurement apparatus, as they persist in measurements without a resonator. 
\begin{table}[b]
	\caption{\label{tab:NEMSModes}
		Resonance frequencies $f$, Q-factors, and effective masses $m_{\textrm{eff}}$ of the trampoline resonator calculated from FEM simulations.}
	\begin{ruledtabular}
		\begin{tabular}{cccc}
			Mode No. & $f_i [\textrm{kHz}]$ & $Q_i$ & $m_{\textrm{eff},i} [\textrm{kg}]$ \\
			\hline
			1& 23.05 & 110000 & 4.52e-12 \\
			2& 49.29 & -      & 7.99e-23 \\
			3& 49.32 & -      & 9.49e-24 \\
			4& 49.37 & -      & 7.14e-21 \\
			5& 68.02 & 150000 & 6.06e-13 \\
			\vdots& \vdots & \vdots & \vdots\\
			13& 114.05& 112000 & 2.23e-13\\
		\end{tabular}
	\end{ruledtabular}
\end{table}

For the mathematical model of the nanomechanical resonator, we assume that the feedback control holds the resonator in a weakly excited state, and therefore, nonlinear effects \cite{schmid2016fundamentals} due to, e.g., geometry, actuation, detection, or damping can be neglected. Thus, the dynamics of the resonator can be decomposed into an ensemble of uncoupled resonant modes of the form
\begin{equation} \label{eqn:spring_mass_ode}
    m_{\textrm{eff},i}\ddot{z}_i + \gamma_{\textrm{eff},i}\dot{z}_i + k_{\textrm{eff},i}z_i  = F_i \textrm{,}
\end{equation}
with $z_i$ being the displacement and $F_i$ the cumulated external force. Moreover, $m_{\textrm{eff},i}$, $\gamma_{\textrm{eff},i}$, and $k_{\textrm{eff},i}$ represent the effective mass, the effective damping, and the effective spring constant of the resonator \cite{schmid2016fundamentals}. The force $F_i(t)$ includes an unknown random force due to the disturbance $\eta_i(t)$ from, for example, gas collision or thermomechanical fluctuations, and a second deterministic part due to the actuator input $u(t)$. Since the system is placed in a vacuum, the dominant source of noise is caused by thermomechanical fluctuations in the material described by the fluctuation dissipation theorem \cite{kubo1966fluctuation, callen1951irreversibility}. The resulting force noise has a one-sided power spectral density (PSD) with units of [N$^2$/Hz] of, 
\begin{equation}
	S_{\eta\eta,i} = 4k_{\textrm{B}}T\gamma_{\textrm{eff},i}
\end{equation} 
where $T$ is the temperature and $k_\textrm{B}$ is the Boltzmann constant. 
Rewriting (\ref{eqn:spring_mass_ode}) in form of (\ref{eqn:LTI1}) for the $i$-th mode  and defining $\Omega_i =\sqrt{\frac{k_{\textrm{eff},i}}{m_{\textrm{eff,}i}}}$, $Q_i=\frac{\Omega_i m_{\textrm{eff,}i}}{\gamma_{\textrm{eff,}i}}$, and the force $F = \sqrt{S_{\eta\eta,i}}\eta_i + b_{\textrm{F},i} u$ yields
\begin{eqnarray} \label{eqn:singleModeSS}
      \frac{\mathrm{d}}{\mathrm{d}t}
      \mathbf{x}_i
      =
      \underbrace{
      \begin{bmatrix}
          0   &  1\\
          -\Omega_i^2   &  -\frac{\Omega_i}{Q_i}
      \end{bmatrix}}_{\mathbf{A}_i}
      \mathbf{x}_i
      +
      \underbrace{
      \begin{bmatrix}
          0  \\ \frac{b_{\textrm{F},i}}{m_{\textrm{eff,}i}}
      \end{bmatrix}}_{\mathbf{B}_i} u
      +
      \underbrace{
      \begin{bmatrix}
          0 \\
          \sqrt{S_{\eta\eta,i}}
      \end{bmatrix}}_{\mathbf{G}_i}\eta_i
\end{eqnarray}
with the state $\mathbf{x}_i = \transpose{\begin{bmatrix} z_i & v_i \end{bmatrix}}$ where $v_i = \dot{z}_i$. Note that the effective mass $m_{\textrm{eff},i}$ can be calculated directly from the shape of the corresponding $i$-th eigenmode, see App.~\ref{app:effective_mass}. Moreover, a calibration procedure according to App.~\ref{app:force_calibration} was applied to determine the coefficients $b_{\textrm{F},i}$.

Since nanomechanical systems have infinite oscillation modes \cite{schmid2016fundamentals}, we must truncate our model to some finite number for it to be computable. Without any restrictions of generality, the nanomechanical system of Fig.~\ref{fig:FIG1}(\subref{fig:resonator_picture}) is modeled considering the first three out-of-plane modes; see Fig.~\ref{fig:FIG1}(\subref{fig:ResSpec}) for the corresponding modal shapes. This results in the following block-diagonal system structure
\begin{subequations}\label{eqn:MultharmSS}
    \begin{equation} \label{eqn:truncSSNEMS}
    \begin{array}{ccccc}
            \frac{\mathrm{d}}{\mathrm{d}t}
            \begin{bmatrix}
            	\mathbf{x}_1 \\ \mathbf{x}_2 \\ \mathbf{x}_3
           	\end{bmatrix}
            &=&
            \begin{bmatrix}
                \mathbf{A}_1   &  0 & 0\\
                0   &  \mathbf{A}_2 & 0 \\
                0   &  0    & \mathbf{A}_3
            \end{bmatrix}
            \begin{bmatrix}
            	\mathbf{x}_1 \\ \mathbf{x}_2 \\ \mathbf{x}_3
            \end{bmatrix}
            &+&
            \begin{bmatrix}
                \mathbf{B}_{1} \\ \mathbf{B}_{2} \\ \mathbf{B}_{3} 
            \end{bmatrix}u
           	\\
        	&&&&
        	\\
            &+&
            \begin{bmatrix}
                \mathbf{G}_1   &  0 & 0\\
                0   &  \mathbf{G}_2 & 0 \\
                0   &  0    & \mathbf{G}_3
           \end{bmatrix}
            \begin{bmatrix}
            	\eta_1 \\
            	\eta_2 \\
            	\eta_3
            \end{bmatrix}.
        	&&
        \end{array}
    \end{equation}
% However this model is only able to describe the dynamics of the resonator itself. Other effects which might have additional effect onto the measurement still are not considered. Therefore, it is beneficial to model all information containing processes which influence our measurement. In this regard, most relevant are the oscillation peak at \SI{21}{kHz} and the non-flat noise floor over our frequency range of interest. The peak at \SI{42}{kHz} can be neglected, as its distance to the peaks of interest in far enough. Likewise, the noise for frequency lower than \SI{10}{kHz} can be neglected. To model the two described processes we introduce another state space model of the form
A laser-Doppler vibrometer measures the trampoline's velocity, which corresponds to the sum of the considered mode displacements, i.e., $y_n(t) = \sum_{i=1}^{3}\dot{z}_i$. Thus, for the considered nanomechanical system, the output equation reads as 
    \begin{equation}
        y_n = 
        \begin{bmatrix}
            \mathbf{C}_1 & \mathbf{C}_2 & \mathbf{C}_3
        \end{bmatrix}
        \begin{bmatrix}
            \mathbf{x}_1 \\
            \mathbf{x}_2 \\
            \mathbf{x}_3
        \end{bmatrix}
    \end{equation}
\end{subequations}
with $\mathbf{C}_i=[0\quad1]$.

The mathematical model (\ref{eqn:MultharmSS}) only covers the nominal behavior of the nanomechanical system depicted in Fig.~\ref{fig:FIG1}(\subref{fig:resonator_picture}). Therefore, we will compare the model with measurements in the next step to assess the model quality. Figure~\ref{fig:FIG1}(\subref{fig:ResSpec}) depicts the PSD of the measurement signal $y(t)$ from \SI{0}{kHz} to \SI{180}{kHz}. The spectrum shows prominent peaks at around \SI{21}{kHz}, \SI{23}{kHz}, \SI{43}{kHz}, \SI{68}{kHz}, \SI{114}{kHz}, and \SI{163}{kHz}. Finite element simulations reveal that the \SI{23}{kHz}, \SI{68}{kHz}, \SI{114}{kHz} and \SI{168}{kHz} peaks correspond to the resonator's first four out-of-plane oscillation modes. Fig.~\ref{fig:FIG1}(\subref{fig:ResSpec}) also depicts the first three mode shapes. Table~\ref{tab:NEMSModes} summarizes the corresponding resonance frequencies $f$, the Q-factors, and the effective masses $m_{\textrm{eff}}$ of the first resonator modes from FEM simulations. Only modes predominantly oscillating out of plane significantly contribute to the resonator's effective mass $m_{\textrm{eff}}$ and, therefore, to the overall deformation energy of the resonator.

As Fig.~\ref{fig:FIG1}(\subref{fig:ResSpec}) shows, the model (\ref{eqn:MultharmSS}) does not cover the measured peaks at \SI{21}{kHz} and \SI{42}{kHz}. They likely originate from the measurement apparatus, as they persist in the measurements without the resonator. Moreover, the model (\ref{eqn:MultharmSS}) does not include the slow increase in the PSD from \SI{10}{kHz} onward and the characteristics below  \SI{10}{kHz}. In the following, the peak at \SI{42}{kHz} and the low-frequency characteristic will be neglected because they only have a minor influence on the measured output signal $y(t)$, which is dominated by the first three modes. In contrast, the peak at \SI{21}{kHz} and the slow increase in the PSD from \SI{10}{kHz} onward are modeled as a disturbance model, which in state-space form reads as
\begin{subequations} \label{eqn:DistSS}
	\begin{equation} \label{eqn:noise_sys}
		\frac{\mathrm{d}}{\mathrm{d}t}
		\begin{bmatrix}
			\mathbf{x}_\textrm{np} \\
			\mathbf{x}_\textrm{nf}
		\end{bmatrix}
		=
		\begin{bmatrix}
			\mathbf{A}_\textrm{np} & 0 \\
			0 & \mathbf{A}_\textrm{nf} 
		\end{bmatrix}
		\begin{bmatrix}
			\mathbf{x}_\textrm{np} \\
			\mathbf{x}_\textrm{nf}
		\end{bmatrix}
		+
		\begin{bmatrix}
			\mathbf{G}_\textrm{np} & 0\\
			0 & \mathbf{G}_\textrm{nf}\\
		\end{bmatrix}
		\begin{bmatrix}
			\mathbf{\eta}_\textrm{np} \\
			\mathbf{\eta}_\textrm{nf}
		\end{bmatrix}
	\end{equation}
 	\begin{equation} \label{eqn:truncMeasNEMS}
		y_d =
		\begin{bmatrix}
		\mathbf{C}_\textrm{np} & \mathbf{C}_\textrm{nf}
		\end{bmatrix}
		\begin{bmatrix}
			\mathbf{x}_\textrm{np} \\ \mathbf{x}_\textrm{nf}
		\end{bmatrix} + \nu.
	\end{equation}
\end{subequations}
Here, the subsystem ($\mathbf{A}_{\textrm{np}}, \mathbf{G}_{\textrm{np}}$) is a weakly damped oscillator with a resonance frequency of \SI{21}{kHz}. The subsystem ($\mathbf{A}_{\textrm{nf}}, \mathbf{G}_{\textrm{nf}}$) corresponds to a band-pass filter with a lower cutoff frequency at \SI{10}{kHz} and a higher cutoff frequency at \SI{200}{kHz}, reflecting the increase in the PSD from \SI{10}{kHz} onward. Moreover, $\eta_{\textrm{np}}$ and $\eta_{\textrm{nf}}$ are white Gaussian measurement noise processes driving the system. The PSD of the output $y_d(t)$ of the disturbance model (\ref{eqn:DistSS}) is depicted as the red dashed line in Fig.~\ref{fig:FIG1}(\subref{fig:ResSpec}). 

Combining the nominal  model (\ref{eqn:MultharmSS}) and the disturbance model (\ref{eqn:DistSS}) leads to a state-space description of the measured behavior in the form \eqref{eqn:LTI1} and \eqref{eqn:LTI2} with the extended state $\mathbf{x} = \begin{bmatrix}\mathbf{x}_1 & \mathbf{x}_2 & \mathbf{x}_3 & \mathbf{x}_{\textrm{np}} & \mathbf{x}_{\textrm{nf}} \end{bmatrix}$ and the output $y=y_n+y_d$, resulting in the matrices
\begin{equation*}
    \begin{array}{cc}
        \underbrace{
        \begin{bmatrix}
            \mathbf{A}_1 & 0 & 0 & 0 & 0 \\
            0 & \mathbf{A}_2 & 0 & 0 & 0 \\
            0 & 0 & \mathbf{A}_3 & 0 & 0 \\
            0 & 0 & 0 & \mathbf{A}_{\textrm{np}} & 0 \\
            0 & 0 & 0 & 0 & \mathbf{A}_{\textrm{nf}}
        \end{bmatrix}\text{,}
        }_{:=\mathbf{A}}
        &
        \underbrace{
        \begin{bmatrix}
            \mathbf{B}_1 \\
            \mathbf{B}_2 \\
            \mathbf{B}_3 \\
            0 \\
            0
        \end{bmatrix}\text{,}
        }_{:=\mathbf{B}}
        \\
        \\
        \underbrace{
        \begin{bmatrix}
            \mathbf{G}_1 & 0 & 0 & 0 & 0 \\
            0 & \mathbf{G}_2 & 0 & 0 & 0 \\
            0 & 0 & \mathbf{G}_3 & 0 & 0 \\
            0 & 0 & 0 & \mathbf{G}_{\textrm{np}} & 0 \\
            0 & 0 & 0 & 0 & \mathbf{G}_{\textrm{nf}}
        \end{bmatrix}\text{,}
        }_{:=\mathbf{G}}
         &
         \underbrace{
         \begin{bmatrix}
             \mathbf{C}_1 & \mathbf{C}_2 & \mathbf{C}_3 & \mathbf{C}_{\textrm{np}} & \mathbf{C}_{\textrm{nf}}
         \end{bmatrix}\text{.}
         }_{:=\mathbf{C}}
    \end{array}
\end{equation*}

\subsection{Optimal Estimation} \label{sec:est}
Assuming that the system \eqref{eqn:LTI1} and \eqref{eqn:LTI2} is fully observable (see App.~\ref{app:observe_control}), we can reconstruct the states of the nanomechanical system from the given output measurements $\mathbf{y}$. This is always the case for the matrices above. Therefore, we introduce a new dynamic system called state observer for the state estimate $\mathbf{x}_{\textrm{f}}$ that should asymptotically converge to the true state $\mathbf{x}$ of \eqref{eqn:LTI1} for $t\rightarrow \infty$. This reads as \cite{luenberger1964observing}
\begin{equation} \label{eqn:estimator}
    \frac{\mathrm{d}}{\mathrm{d}t}\mathbf{x}_{\textrm{f}}= \mathbf{A}\mathbf{x}_{\textrm{f}} + \mathbf{B}\mathbf{u} + \mathbf{K}_{\textrm{f}} (\mathbf{y} - \mathbf{C}\mathbf{x}_{\textrm{f}}) \textrm{,}
\end{equation}
with the initial condition $\mathbf{x}_{\textrm{f}}(0) = \mathbf{x}_{{\textrm{f}},0}$ and the observer gain matrix $\mathbf{K}_{\textrm{f}}$ \cite{luenberger1964observing}. 
Suppose $\mathbf{K}_{\textrm{f}}$ is chosen as
\begin{subequations}\label{eqn:Klamfeqn}
	\begin{equation}
	    \mathbf{K}_{\textrm{f}}=\mathbf{\Sigma}_{\textrm{f}} \transpose{\mathbf{C}}\mathbf{R}^{-1}
	\end{equation}
 with a positive definite $\boldsymbol{\Sigma_{\textrm{f}}}$ according to
    \begin{equation}
        \frac{\mathrm{d}}{\mathrm{d}t}\mathbf{\Sigma}_{\textrm{f}}=
        \mathbf{A}\mathbf{\Sigma}_{\textrm{f}} + \mathbf{\Sigma}_{\textrm{f}}\transpose{\mathbf{A}}+\mathbf{G}\transpose{\mathbf{G}}
        -\mathbf{\Sigma}_{\textrm{f}} \transpose{\mathbf{C}}\mathbf{R}^{-1}\mathbf{C}\mathbf{\Sigma}_{\textrm{f}} \label{eqn:MatrixRiccati}.
    \end{equation}
\end{subequations}
% \begin{subequations}\label{eqn:Klamfeqn}
% 	\begin{eqnarray}
% 		\mathbf{K}_{\textrm{f}}&=&\mathbf{\Sigma}_{\textrm{f}} \transpose{\mathbf{C}}\mathbf{R}^{-1}\\
% 		\frac{\mathrm{d}}{\mathrm{d}t}\mathbf{\Sigma}_{\textrm{f}}&=&
% 			\mathbf{A}\mathbf{\Sigma}_{\textrm{f}} + \mathbf{\Sigma}_{\textrm{f}}\transpose{\mathbf{A}}+\mathbf{G}\transpose{\mathbf{G}}
% 			  -\mathbf{\Sigma}_{\textrm{f}} \transpose{\mathbf{C}}\mathbf{R}^{-1}\mathbf{C}\mathbf{\Sigma}_{\textrm{f}} \label{eqn:MatrixRiccati}.
% 	\end{eqnarray}   
% \end{subequations} 
In that case, the state observer (\ref{eqn:Klamfeqn}) minimizes the expected mean square error $\mathbb{E}[||\mathbf{x}-\mathbf{x}_{\textrm{f}}||_2]$ for Gaussian white noise processes $\eta$ and $\nu$ according to (3) with $\boldsymbol{\Sigma}_{\textrm{f}}$ as the covariance matrix of the estimation error. The resulting optimal state estimator (\ref{eqn:estimator}) and (\ref{eqn:Klamfeqn}) is the well-known Kalman-Bucy filter \cite{kalmBu}.
The differential matrix Riccati equation (\ref{eqn:MatrixRiccati}) describes the time evolution of the covariance of the estimation error. Choosing a positive definite initial condition $\mathbf{\Sigma}_{\textrm{f}}(0)=\mathbf{\Sigma}_{\textrm{f},0}$ ensures (\ref{eqn:Klamfeqn}) is well-posed, and the solution $\mathbf{\Sigma}_{\textrm{f}}(t)$ remains positive definite for all $t > 0$ \cite{kalmBu}. Moreover, a closer look at (\ref{eqn:MatrixRiccati}) reveals that due to \eqref{eqn:LTI1} and \eqref{eqn:LTI2} being fully observable, the differential matrix Riccati equation (\ref{eqn:MatrixRiccati}) converges to a unique stationary point for a positive definite initial condition $\mathbf{\Sigma}_{\textrm{f},0}$. Therefore, the observer gain matrix $\mathbf{K}_{\textrm{f}}$ also converges to some steady-state value.

\subsection{Optimal Smoothing} \label{sec:smooth}
The Kalman-Bucy filter determines the optimal estimate $\mathbf{x}_{\textrm{f}}(t)$ for a given time sequence of measurement $\mathbf{y}(t')$ with $t' \leq t$, i.e., it is a \emph{causal} estimator and can be implemented in real-time. For many applications in signal processing, this causality restriction is not necessary if information in the form of future measurements is available, e.g., if data is recorded and then post-processed. So-called \emph{smoothers}, such as the RTS-smoother proposed in \cite{rtsSmo}, exploit this information by conditioning a smoothed state estimate $\mathbf{x}_{\textrm{s}}$ on past and future measurements of the output $\mathbf{y}$. 
This is achieved by filtering the data forward in time by applying the Kalman-Bucy filter (\ref{eqn:estimator}) and (\ref{eqn:Klamfeqn}) and then improving the estimate by backwards filtering. The time-reversed dynamics are given by 
\begin{equation} \label{eqn:smootherStateEstimate}
    \frac{\mathrm{d}}{\mathrm{d}t}\mathbf{x}_{\textrm{s}} = \mathbf{A}\mathbf{x}_{\textrm{s}}+\mathbf{G}\transpose{\mathbf{G}}\mathbf{\Sigma}_{\textrm{f}}^{-1}(\mathbf{x}_{\textrm{s}} - \mathbf{x}_{\textrm{f}}),
\end{equation}
and the covariance matrix $\mathbf{\Sigma}_{\textrm{s}}$ of the smoother's error $\mathbf{x}_{\textrm{s}} - \mathbf{x}$ results from  
\begin{equation} \label{eqn:SmootherEqn}
    \frac{\mathrm{d}}{\mathrm{d}t}\mathbf{\Sigma}_{\textrm{s}}=
    (\mathbf{A}+\mathbf{G}\transpose{\mathbf{G}}\mathbf{\Sigma}_{\textrm{f}}^{-1})\mathbf{\Sigma}_{\textrm{s}}+\mathbf{\Sigma}_{\textrm{s}}\transpose{(\mathbf{A}+\mathbf{G}\transpose{\mathbf{G}}\mathbf{\Sigma}_{\textrm{f}}^{-1})}-\mathbf{G}\transpose{\mathbf{G}},
\end{equation}
starting from a given filtered state estimate $\mathbf{x}_{\textrm{f}}(t)$ and error covariance matrix $\mathbf{\Sigma}_{\textrm{f}}(t)$ for $0 \leq t \leq T$ due to \eqref{eqn:estimator} and \eqref{eqn:Klamfeqn}. The smoother equations \eqref{eqn:smootherStateEstimate} and \eqref{eqn:SmootherEqn} are solved backward in time from the final conditions $\mathbf{x}_{\textrm{s}}(T) = \mathbf{x}_{\textrm{f}}(T)$ and $\mathbf{\Sigma}_{\textrm{s}}(T) = \mathbf{\Sigma}_{\textrm{f}}(T)$.
This way, one obtains the maximum-likelihood state estimate conditioned on all measurements in the fixed time interval $0 \leq t \leq T$ under the assumed system dynamics \cite{fraser1969optimum, rtsSmo, mayne1966solution}. 

\subsection{Optimal Control} \label{sec:opt_cntrl}
We want to find an optimal state feedback law for some cost function. While this is a challenging problem in general, a unique closed-form solution exists for linear stochastic system dynamics with a quadratic cost function.
\begin{equation}
    J = \lim_{t_{\textrm{f}}\to \infty}\frac{1}{2t_{\textrm{f}}}\int_{-t_{\textrm{f}}}^{t_{\textrm{f}}}\transpose{\mathbf{x}}\mathbf{M}\mathbf{x} + \transpose{\mathbf{u}}\mathbf{N}\mathbf{u}\mathrm{d}t,
\end{equation}
with $\mathbf{M}$ and $\mathbf{N}$ being symmetric and positive definite. 
In the steady-state case, the solution to this so-called (infinite-horizon) stochastic linear quadratic regulator (LQR) problem is given by the linear state feedback law with feedback gain $\mathbf{K}_\textrm{c}$, i.e.,
\begin{equation} \label{eqn:LQR_control_law}
    \mathbf{u} = -\mathbf{K}_{\textrm{c}} \mathbf{x} = -\mathbf{N}^{-1}\transpose{\mathbf{B}}\mathbf{V}\mathbf{x}
\end{equation}
where the matrix $\mathbf{V}$ (which represents the value function) is given by the solution of the algebraic Riccati equation
\begin{equation} \label{eqn:lqrRaccati}
    \mathbf{0} = \mathbf{V}\mathbf{A} +\transpose{\mathbf{A}}\mathbf{V} - \mathbf{V}\mathbf{B}\mathbf{N}^{-1}\transpose{\textbf{B}}\mathbf{V} + \mathbf{M}.
\end{equation}
A solution of \eqref{eqn:lqrRaccati} exists if the system \eqref{eqn:LTI1} is stabilizable (see App.~\ref{app:observe_control}), which is always the case for the resonator structure in Sec.~\ref{sec:modeling} if $b_{\textrm{F},i} \neq 0$.
The combination of the Kalman-Bucy filter (\ref{eqn:estimator}) and (\ref{eqn:Klamfeqn}) and the LQR (\ref{eqn:LQR_control_law}) and (\ref{eqn:lqrRaccati}), i.e., replacing $\mathbf{x}$ in \eqref{eqn:LQR_control_law} by the estimated state $\mathbf{x}_{\textrm{f}}$, yields the optimal output feedback for the system (\ref{eqn:LTI1}) with Gaussian white noise processes $\boldsymbol{\eta}$ and $\boldsymbol{\nu}$ according to \eqref{eqn:noises}, also known as \emph{linear quadratic Gaussian regulator} (LQG) \cite{athans1971role}.

\section{Optimal Estimation of Momentum Kicks} \label{chap:ImpEst}
Using the framework summarized in the previous section, we now want to estimate the unknown momentum kick provided to the (feedback-controlled) nanomechanical system in (\ref{eqn:MultharmSS}) and (\ref{eqn:DistSS}), assuming a single idealized Dirac-like disturbance at a point in time $t_p \in [t_a,t_b]$ for a measured time trace of input and output data $\boldsymbol{\mathcal{D}}=\left[ \mathbf{y}(t), \mathbf{u}(t) \right]_{t_a \leq t \leq t_b}$ coming from our nanomechanical system.

The momentum kick at $t_p$ will instantaneously change the motional state $\mathbf{x}$ of the resonator. This discontinuous change in the dynamic variables of our system will break parts of the temporal correlation between time points before and after the impulse. Hence, we split our time trace into two intervals $\boldsymbol{\mathcal{D}}_1= [\mathbf{y}(t),\mathbf{u}(t)]_{t_a\leq t<t_p}$ and $\boldsymbol{\mathcal{D}}_2= [\mathbf{y}(t),\mathbf{u}(t)]_{t_p\leq t \leq t_b}$, i.e., before and after the kick, both of which are accurately described by the stochastic model \eqref{eqn:LTI1} and \eqref{eqn:LTI2}. We can then estimate the change of the motional state $\Delta\mathbf{x}(t_p)$ due to the momentum kick by 
\begin{equation} \label{eq:kick_estimate}
    \Delta\hat{\mathbf{x}}(t_p) = \hat{\mathbf{x}}(t_p)|\boldsymbol{\mathcal{D}}_1 - \hat{\mathbf{x}}(t_p)|\boldsymbol{\mathcal{D}}_2 \textrm{,}
\end{equation}
where $\Delta\hat{\mathbf{x}}(t_p)$ denotes the optimal estimate of $\Delta\mathbf{x}$ at $t_p$ conditioned on data $\boldsymbol{\mathcal{D}}_{1}$ and $\boldsymbol{\mathcal{D}}_{2}$, respectively. 

Since $\hat{\mathbf{x}}(t_p)|\boldsymbol{\mathcal{D}}_1$ is, per definition, only conditioned on past data, it is directly given by solving the Kalman-Bucy filter equations (\ref{eqn:estimator}) and (\ref{eqn:Klamfeqn}), yielding $\mathbf{x}_{{\textrm{f}},\boldsymbol{\mathcal{D}}_1}(t)$ and $\boldsymbol{\Sigma}_{\textrm{f},\boldsymbol{\mathcal{D}}_1}(t)$ for $t \in [t_a,t_p)$ and $\hat{\mathbf{x}}(t_p)|\boldsymbol{\mathcal{D}}_1 = \lim_{t\to t_p} \mathbf{x}_{\textrm{f},\boldsymbol{\mathcal{D}}_1}(t_p)$. 
Since the Kalman-Bucy filter obeys a differential equation, its estimates depend on the initial states $\mathbf{x}_{\textrm{f}}(t_a)$ and $\boldsymbol{\Sigma}_{\textrm{f}}(t_a)$. The final estimate becomes independent of the initial value for sufficiently long time traces where (\ref{eqn:MatrixRiccati}) becomes stationary.

Conversely, $\hat{\mathbf{x}}(t_p)|\boldsymbol{\mathcal{D}}_2$ is only conditioned on future data, which is given by solving the RTS-smoother equations \eqref{eqn:smootherStateEstimate} and \eqref{eqn:SmootherEqn} for $t \in [t_p,t_b]$, which yields $\mathbf{x}_{\textrm{s},\boldsymbol{\mathcal{D}}_2}(t)$ and $\boldsymbol{\Sigma}_{\textrm{s},\boldsymbol{\mathcal{D}}_2}(t)$ for $t \in [t_p,t_b]$ and $\hat{\mathbf{x}}(t_p)|\boldsymbol{\mathcal{D}}_2 = \mathbf{x}_{\textrm{s},\boldsymbol{\mathcal{D}}_2}(t_p)$. This entails solving the Kalman-Bucy filter equations forward in time from initial conditions $\mathbf{x}_{\textrm{f},\boldsymbol{\mathcal{D}}_2}(t_p)$ and $\boldsymbol{\Sigma}_{\textrm{f},\boldsymbol{\mathcal{D}}_2}(t_p)$. 
Assuming that the unknown momentum kick is random with zero mean, the expectation value of the state estimates before and after the kick does not change. Similarly, the uncertainty and, thus, the variance of the position estimates will remain unchanged. In contrast, the variance of the velocity estimates will increase, i.e., 
\begin{subequations}
    \begin{equation}
        \mathbf{x}_{\textrm{f},\boldsymbol{\mathcal{D}}_2}(t_p) = \mathbf{x}_{\textrm{f},\boldsymbol{\mathcal{D}}_1}(t_p) \text{,}
    \end{equation}
    \begin{equation}
        \boldsymbol{\Sigma}_{\textrm{f},\boldsymbol{\mathcal{D}}_2}(t_p) = \boldsymbol{\Sigma}_{\textrm{f},\boldsymbol{\mathcal{D}}_1}(t_p) + \boldsymbol{\Sigma}_{\sigma} \text{,}
    \end{equation}
    \begin{equation}
        \boldsymbol{\Sigma}_{\sigma} = \text{diag}\big(\left[0 \quad \sigma_{p,1}^2 \quad \cdots \quad 0 \quad \sigma^2_{p,3} \quad 0 \quad \cdots \quad 0 \right]\big) \text{,}
    \end{equation}
\end{subequations}
where $\boldsymbol{\Sigma}_{\sigma}$ represents the expected increase in uncertainty according to the assumptions on the momentum kicks. In scenarios with strong expected kicks, the matrix $\boldsymbol{\Sigma}_{\sigma}$ adds values $\sigma^2_{p,i} \forall i \in \{1,2,3\}$ several orders of magnitude higher than the steady-state entries of $\boldsymbol{\Sigma}_{\textrm{f},\boldsymbol{\mathcal{D}}_1}(t_p)$ to the diagonal entries of $\boldsymbol{\Sigma}_{\textrm{f},\boldsymbol{\mathcal{D}}_2}(t_p)$, which effectively implies that the filter is uninformed about the initial velocity after the kick. Hence, it is generally advised to overestimate values of $\sigma^2_{p,i}$.

\begin{figure}[h!]
    \begin{subfigure}{0.45\textwidth}
        \caption{}
        \centering
        \def\svgwidth{\textwidth}
        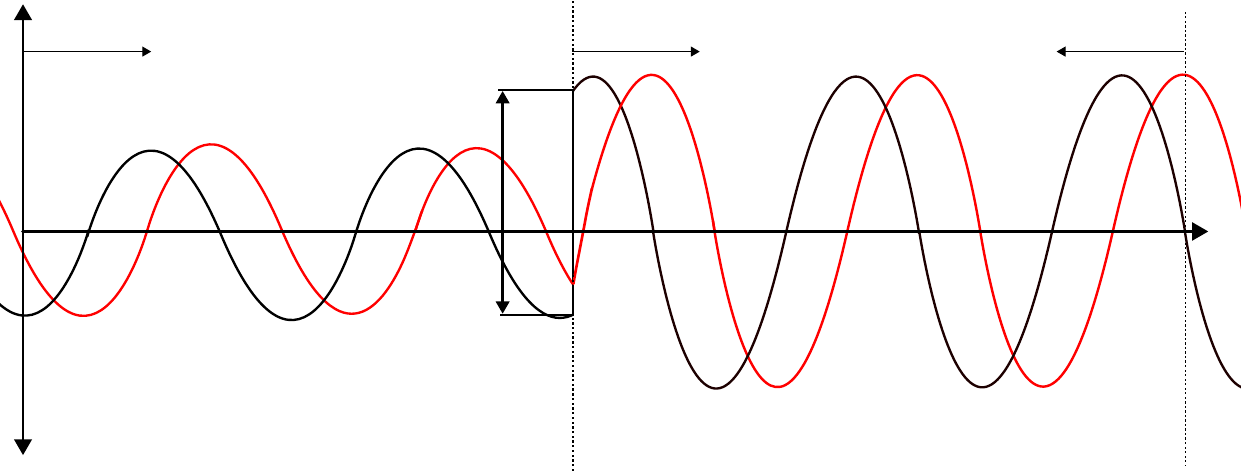
        \label{fig:impSine}
    \end{subfigure}
    \begin{subfigure}{0.45\textwidth}
        \caption{}
        \centering
        \def\svgwidth{\textwidth}
        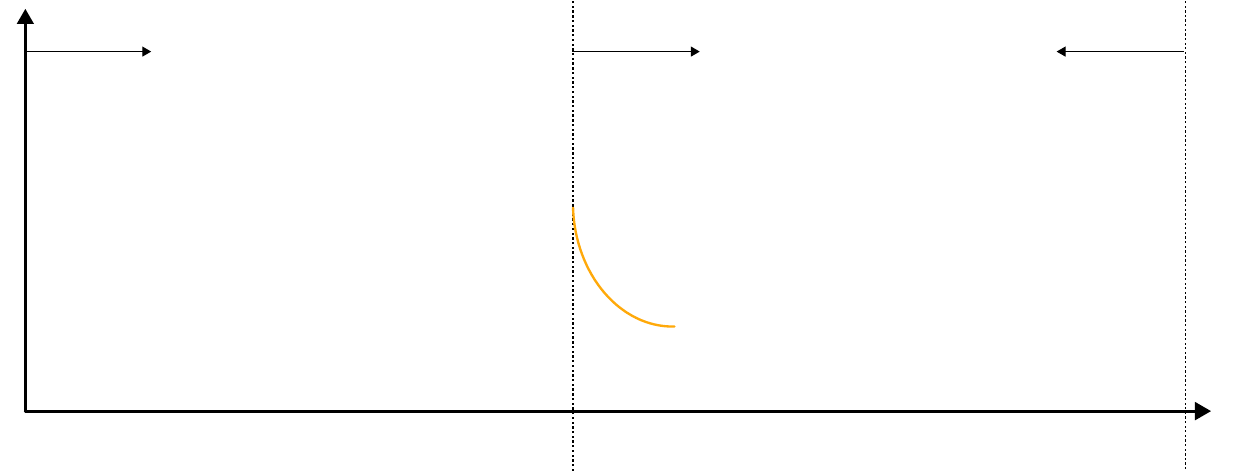
        \label{fig:impCov} 
    \end{subfigure}
\caption{Schematic overview of the complete estimation process: Figure (\subref{fig:impSine}) shows the estimated time traces before and after the kick at $t=t_p$, which results in a strong discontinuity in velocity that can be retrieved by applying Kalman-Bucy filtering to the data before the kick and RTS-smoothing to the data after the kick. Figure (\subref{fig:impCov}) shows the corresponding time evolution of the velocity entries of the covariance matrices.}
\label{fig:ImpFilter}
\end{figure}

An illustration of the complete procedure applied to a time trace of a single-mode resonator is given in Figure \ref{fig:ImpFilter}.
The resulting estimate $\Delta\hat{\mathbf{x}}(t_p)$ given by \eqref{eq:kick_estimate} represents the effect of the momentum kick extracted from the time traces, where the change in position $\Delta z$ is expected to be close to zero change in velocity $\Delta v$ is directly proportional to the momentum transferred onto the resonator.
The covariance of the estimate is bounded by
\begin{equation} \label{eqn:upSensbound}
    \mathbf{\Sigma}_{\Delta\hat{\mathbf{x}}}(t_p) \preccurlyeq \mathbf{\Sigma}_{\textrm{f},\boldsymbol{\mathcal{D}}_1}(t_p) + \mathbf{\Sigma}_{\textrm{s},\boldsymbol{\mathcal{D}}_2}(t_p)\text{,}
\end{equation}
as the two estimates are generally correlated.

\section{Results}
\begin{figure*}[hbt!]
    \centering
    \begin{subfigure}[c]{1\textwidth}
        \caption{}
        % This file was created by matlab2tikz.
%
%The latest updates can be retrieved from
%  http://www.mathworks.com/matlabcentral/fileexchange/22022-matlab2tikz-matlab2tikz
%where you can also make suggestions and rate matlab2tikz.
%
\definecolor{mycolor1}{rgb}{0.00000,0.44700,0.74100}%
\definecolor{mycolor2}{rgb}{0.85000,0.32500,0.09800}%
\definecolor{mycolor3}{rgb}{0.24314,0.58824,0.31765}%
\begin{tikzpicture}

\begin{axis}[%
width=1\textwidth,
height=0.33\textwidth,
%at={(2.708in,0.688in)},
%scale only axis,
xmin=1,
xmax=180,
xlabel style={font=\color{white!15!black}},
xlabel={$\omega/2\pi$  (\SI{}{kHz})},
ymode=log,
ymin=1e-18,
ymax=1e-08,
yminorticks=true,
ylabel style={font=\color{white!15!black}},
ylabel={$S_{yy}(\omega)$ $(\SI{}{m}^2/\SI{}{s}^2\SI{}{Hz})$},
axis background/.style={fill=white},
legend style={legend cell align=left, align=left, draw=white!15!black, at={(1,0)}, anchor= south east}
]
%\addplot [color=mycolor1, line width=1.0pt, forget plot]
\addplot [color=mycolor1, forget plot]
  table[]{figures/Data/spec-1.tsv};
%\addplot [color=mycolor1, line width=1.0pt]
\addplot [color=mycolor1]
  table[]{figures/Data/spec-2.tsv};
\addlegendentry{uncontrolled}

\addplot [color=mycolor2, forget plot]
  table[]{figures/Data/spec-3.tsv};
\addplot [color=mycolor2]
  table[]{figures/Data/spec-4.tsv};
\addlegendentry{controlled}

\addplot [color=mycolor3, forget plot]
  table[]{figures/Data/spec-5.tsv};
\addplot [color=mycolor3]
  table[]{figures/Data/spec-6.tsv};
\addlegendentry{innovation}

\end{axis}

\end{tikzpicture}%
        \label{fig:SpecCon}
    \end{subfigure}
    \hfill
    \begin{subfigure}[h]{0.49\textwidth}
        %\centering
        \caption{}
        % This file was created by matlab2tikz.
%
%The latest updates can be retrieved from
%  http://www.mathworks.com/matlabcentral/fileexchange/22022-matlab2tikz-matlab2tikz
%where you can also make suggestions and rate matlab2tikz.
%
\definecolor{mycolor1}{rgb}{0.00000,0.44700,0.74100}%
\definecolor{mycolor2}{rgb}{0.30100,0.74500,0.93300}%
\begin{tikzpicture}

\begin{axis}[%
	%width=0.5\textwidth,
	%height=0.175\textwidth,
	width=1\textwidth,
	height=0.35\textwidth,
	at={(0in,2.8in)},
	%scale only axis,
	xmin=0,
	xmax=200,
	xticklabel=\empty,
	ymin=-1e4,
	ymax=1e4,
	ylabel style={font=\color{white!15!black}},
	ylabel={$y$ (\textmu m/s)},
	axis background/.style={fill=white},
	axis x line*=bottom,
	axis y line*=left
	]
\addplot [color=mycolor1, forget plot]
  table[]{figures/time_traces/time_traces_fb-1.tsv};
\addplot [color=red, forget plot]
  table[]{figures/time_traces/time_traces_fb-2.tsv};
\addplot [color=white!15!black, forget plot]
  table[]{figures/time_traces/time_traces_fb-3.tsv};
\end{axis}

\begin{axis}[%
width=1\textwidth,
height=0.35\textwidth,
at={(0in,2.1in)},
%scale only axis,
xmin=0,
xmax=200,
xticklabel=\empty,
ymin=-45,
ymax=45,
ylabel style={font=\color{white!15!black}},
ylabel={$v_1$ (\textmu m/s)},
axis background/.style={fill=white},
axis x line*=bottom,
axis y line*=left
]

\addplot[area legend, draw=none, fill=mycolor2, forget plot]
table[] {figures/time_traces/time_traces_fb-4.tsv}--cycle;
\addplot [color=red, forget plot]
  table[]{figures/time_traces/time_traces_fb-5.tsv};
\addplot [color=white!15!black, forget plot]
  table[]{figures/time_traces/time_traces_fb-6.tsv};
\end{axis}

\begin{axis}[%
	%width=0.5\textwidth,
	%height=0.175\textwidth,
	width=1\textwidth,
	height=0.35\textwidth,
	at={(0in,1.4in)},
	%scale only axis,
	xmin=0,
	xmax=200,
	xticklabel=\empty,
	ymin=-50,
	ymax=50,
	ylabel style={font=\color{white!15!black}},
	ylabel={$v_2$ (\textmu m/s)},
	axis background/.style={fill=white},
	axis x line*=bottom,
	axis y line*=left
	]

\addplot[area legend, draw=none, fill=mycolor2, forget plot]
table[] {figures/time_traces/time_traces_fb-7.tsv}--cycle;
\addplot [color=red, forget plot]
  table[]{figures/time_traces/time_traces_fb-8.tsv};
\addplot [color=white!15!black, forget plot]
  table[]{figures/time_traces/time_traces_fb-9.tsv};
\end{axis}

\begin{axis}[%
	%width=0.5\textwidth,
	%height=0.175\textwidth,
	width=1\textwidth,
	height=0.35\textwidth,
	at={(0in,0.7in)},
	%scale only axis,
	xmin=0,
	xmax=200,
	xlabel style={font=\color{white!15!black}},
	xlabel={time (\textmu s)},
	ymin=-50,
	ymax=50,
	ylabel style={font=\color{white!15!black}},
	ylabel={$v_3$ (\textmu m/s)},
	axis background/.style={fill=white},
	axis x line*=bottom,
	axis y line*=left
	]

\addplot[area legend, draw=none, fill=mycolor2, forget plot]
table[] {figures/time_traces/time_traces_fb-10.tsv}--cycle;
\addplot [color=red, forget plot]
  table[]{figures/time_traces/time_traces_fb-11.tsv};
\addplot [color=white!15!black, forget plot]
  table[]{figures/time_traces/time_traces_fb-12.tsv};
\addplot [color=white!15!black, forget plot]
  table[]{figures/time_traces/time_traces_fb-13.tsv};
\addplot [color=white!15!black, forget plot]
  table[]{figures/time_traces/time_traces_fb-14.tsv};
\addplot [color=white!15!black, forget plot]
  table[]{figures/time_traces/time_traces_fb-15.tsv};
\end{axis}
\end{tikzpicture}%
        \label{fig:time_traces}
    \end{subfigure}
   \hfill
    \begin{subfigure}[h]{0.49\textwidth}
        \caption{}
        \input{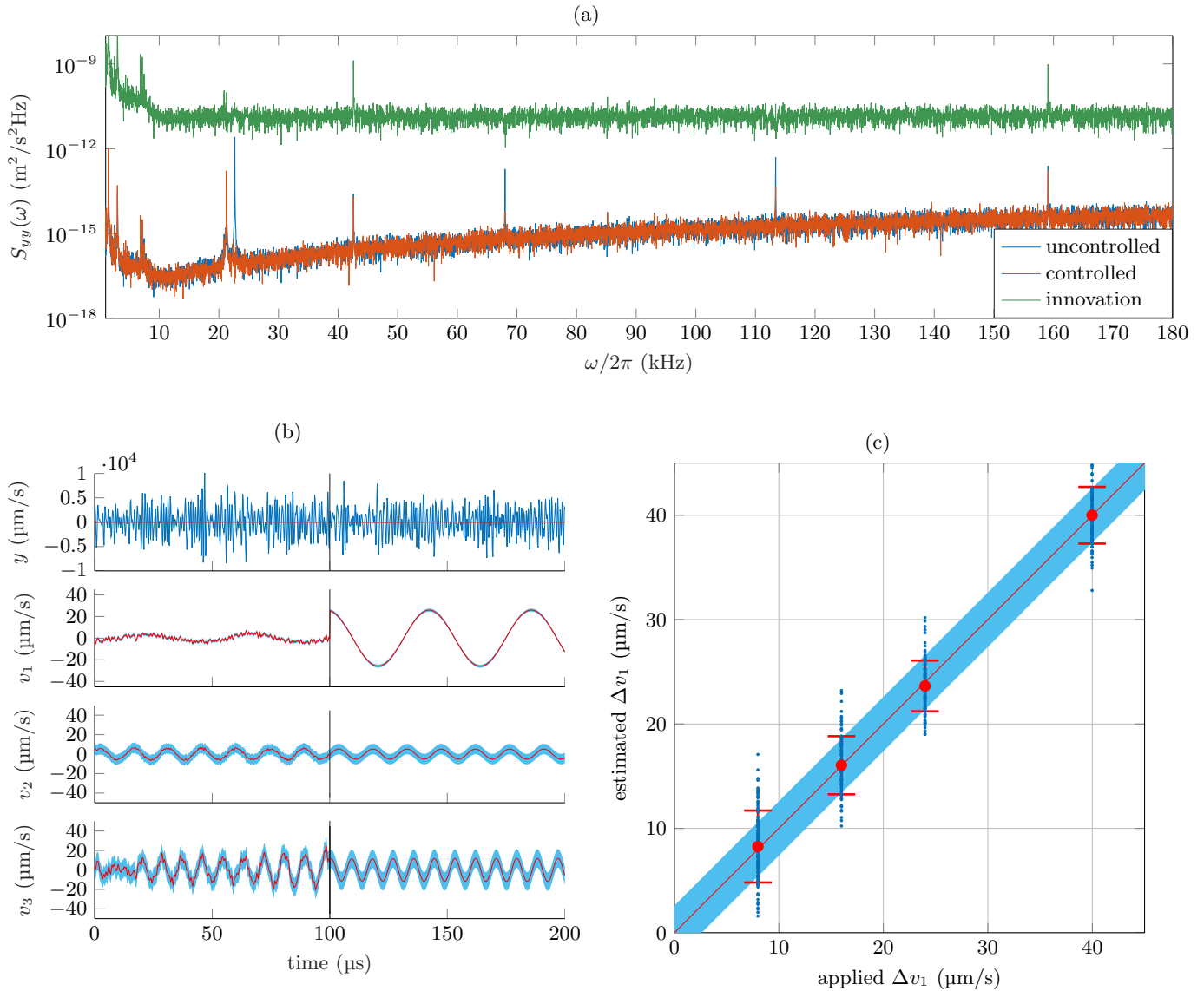}
        \label{fig:ImpReCon}
    \end{subfigure}
    \hfill
    \caption{(\subref{fig:SpecCon}) Measurement spectra with (orange) and without (blue) feedback and the innovation (green).(\subref{fig:time_traces}) Example time trace of a kick applied to the resonator. Top: Measured output signal in blue compared to the estimated output signal in red, buried in noise. Below: Estimated velocities in red of the first three modes. The vertical black line indicates the time of the kick. Uncertainty of the estimated signals (one standard deviation) is shaded in blue. (\subref{fig:ImpReCon}) Statistical relation between applied and estimated kicks. The blue shaded area marks expected uncertainty from theory. The blue dots show individual data points. Red dots and bars show mean and standard deviation of the recorded data.}
    \label{fig:FIG3}
\end{figure*}
To verify that the model and the real-time FPGA implementation of the filter are optimal, we can check the so-called innovation sequence, which is the difference between the estimated output $\mathbf{C}\mathbf{x}_{\textrm{f}}$ and the actual measured output $\mathbf{y}$. If the innovation sequence is a white Gaussian noise process, one can show that the optimal filter was found \cite{bode1950simplified}. The measured innovation of the applied filter is depicted in Figure \ref{fig:FIG3}(\subref{fig:SpecCon}). As can be seen, the innovation shows an almost perfectly flat spectrum, particularly for the frequency range of interest from \SI{10}{kHz} to \SI{130}{kHz}, except for the two unmodeled peaks at \SI{42}{kHz} and \SI{163}{kHz}, far away from our actual oscillator modes.
After activating the controller, the feedback effectively dampens the motion of the three resonator modes, as shown in Figure~\ref{fig:FIG3}(\subref{fig:SpecCon}), practically suppressing the first three modes to the noise floor. 

We artificially inject small momentum kicks with four different magnitudes to the resonator through short voltage impulses $p(t)$ applied to the second actuating gold wire, see Figure \ref{fig:FIG3}(\subref{fig:expSetup1}). Kicks of each magnitude are applied 100 times to obtain statistics on the performance of the proposed estimation algorithm. The magnitudes are chosen between \SI{3.6E-17}{\kilogram \metre \per \second} and \SI{1.8E-16}{\kilogram \metre \per \second}, which is close to the expected accuracy of the estimator. For each time trace, the measurement signal $y(t)$, feedback signal $u(t)$, and voltage pulse $p(t)$ are recorded by the FPGA and transferred to the host computer for post-processing using the optimal estimation algorithm presented in Section~\ref{chap:ImpEst}.

For the chosen small momentum kicks, we only expect significant discontinuities of the velocity of the first mode, $\Delta v_1$, since the forces on the higher modes are several orders of magnitudes smaller.
Considering the effective mass from the FEM simulation (see Table \ref{tab:NEMSModes}), these discontinuities should range between \SI{8E-6}{\metre \per \second} or \SI{3.6E-17}{\kilogram \metre \per \second} and \SI{40E-6}{\metre \per \second} or \SI{1.8E-16}{\kilogram \metre \per \second}. This assumption is confirmed by the exemplary time trace of the measured output $y(t)$ and velocity estimates $v_i$ given in Figure~\ref{fig:FIG3}(\subref{fig:time_traces}), where only $v_1(t)$ shows the expected discontinuity. While the measured output is buried in noise that is orders of magnitude larger than the expected motion of the resonator, the three modal velocities $v_i$ can be estimated with high accuracy.

By comparing the expected discontinuities $\Delta v_1$ with the estimated ones, as shown in Figure~\ref{fig:FIG3}(\subref{fig:ImpReCon}), one can see that the estimates nicely agree with the applied kicks and that there is a clear linear correspondence between these quantities. The standard deviation over the ensemble stays effectively constant at \SI{2.8E-6}{\metre \per \second} or \SI{1.3E-17}{\kilogram \metre \per \second} for test points far away from the precision limit, while a slight increase to \SI{3.3E-6}{\metre \per \second} is noticeable when the measured momentum approaches the practical limit of the current measurement setup at around \SI{6E-6}{\metre \per \second} or \SI{2.7E-17}{\kilogram \metre \per \second}. % when additional effects influencing the precision presumably become dominant.
These empirically observed values are in excellent agreement with the theoretical upper bound according to (\ref{eqn:upSensbound}), which evaluates to \SI{2.9E-6}{\metre \per \second} or \SI{1.3E-17}{\kilogram \metre \per \second} for the given experimental setup. The residual error most likely stems from minor frequency drifts during the measurements and imperfect kicks.

\section{Conclusion and Outlook}
This paper proposes a novel approach to optimally sense momentum kicks applied to a nanomechanical resonator through estimation and control methods. By employing optimal feedback control, we can remove energy transferred to the resonator by consecutive kicks and thus ensure that the resonator operates in the linear regime. Hence, the developed (linear) optimal estimation algorithms can accurately extract the magnitude of each kick from observed stochastic trajectories.
To experimentally validate this approach, we applied the method to a state-of-the-art high-Q nano-mechanical resonator setup. We successfully reconstructed momentum kicks applied to the resonator close to the theoretic limit given by the measurement accuracy and the mathematical model of the resonator.
%Since the proposed method is independent from the used experimental conditions further significant improvements in sensitivity are possible by the usage of oscillators with higher Q-Factors and the implementation in experimental set-ups with higher precision measurements. 

Without further optimization, the current setup can already resolve momentum kicks of approximately \SI{1.3E-17}{\kilogram \metre \per \second}. This is only two orders of magnitude higher than the momentum of particles with a mass of \SI{1}{\kilo\dalton} accelerated to \SI{20}{\kilo\electronvolt}, achievable through standard accelerator stages for mass spectrometry applications.
Recent scientific achievements pushed increasingly massive objects into the quantum regime \cite{rossi2018measurement,delic2020cooling,tebbenjohanns_quantum_2021,xia_motional_2024}. In particular, Heisenberg-limited optical detection, as in \cite{magrini2021real}, would allow for momentum uncertainties around \SI{1E-21}{\kilogram \metre \per \second}, assuming the same mass of the resonator.

However, almost 90\% of the resonator's mass is contributed by the wires required for Lorentz force actuation in the current setup, which could be improved using different methods such as electro-static or capacitive actuation. Combined with the potential of improved measurements and ultra-high-Q resonators using soft-clamping or phononic membrane resonators \cite{tsaturyan2017ultracoherent, engelsen2024ultrahigh}, improving the accuracy by several orders of magnitude seems within reach.
This is remarkable since it indicates that single-molecule mass spectrometry over a large range of particle masses with single \SI{}{\dalton} resolution could be achievable by shooting individual particles at a high-Q resonator and estimating the momentum kick by the proposed method. This is far below the \SI{}{\mega\dalton} range accessible through state-of-the-art nano-mechanical mass spectrometers \cite{stassi_large-scale_2019,sansa2020optomechanical} and compares favorably with other results obtained with significantly smaller devices at cryogenic temperatures \cite{yang_zeptogram-scale_2006}.

\begin{acknowledgments}
The authors would like to thank Johannes Hiesberger for the fabrication of the nanomechanical resonator and Vojtěch Mlynář for support with the FPGA implementation.
\end{acknowledgments}

\appendix
%\section{Appendix}
% \section{Auto-correlation and Power Spectral Density}
% Auto-correlation of a noisy process can be expressed as the long time average of two realizations of that process. In case of a white noise process this results in
% \begin{equation}
% 	\mathbb{E}[\xi(t)\xi(t')] = \lim_{\tilde{t}\to \infty}\frac{1}{2\tilde{t}}\int_{-\tilde{t}}^{\tilde{t}}\xi(t)\xi(t')\mathrm{d}t' = \sigma_\xi^2\delta(t-t')
% \end{equation}
% applying the Fourier transform to the auto-correlation function we get by the Wiener-Khinchi theorem the power spectral density of the white noise process
% \begin{equation}
% 	S_{\xi\xi} = \int_{-\infty}^{\infty}\mathbb{E}[\xi(t)\xi(t')]e^{-j\omega \tau}\mathrm{d}\tau = \sigma_\xi^2
% \end{equation}
% which is constant over all frequencies.

\section{Effective Resonator Mass} \label{app:effective_mass}
The effective mass of a resonator can be calculated directly from the shape of its eigenmodes (see Fig. \ref{fig:FIG1}(\subref{fig:ResSpec}) and its spatial mass density. Assuming the function $\boldsymbol{\phi}_i(\mathbf{r})$ describes the eigenfunction of the $i$-th mode with spatial variable $\mathbf{r}$ and some space-dependent mass density function $\rho(\mathbf{r})$, then the effective mass results in the volume integral \cite{hauer2013general}  
\begin{equation}
	m_{\textrm{eff},i} = \int_V|\boldsymbol{\phi}_i(\mathbf{r})|^2\rho(\mathbf{r})\mathrm{d}V.
\end{equation}

\section{Force calibration} \label{app:force_calibration}
A calibration procedure was applied to determine the coefficients $b_{\textrm{F},i}$ from the mathematical model \eqref{eqn:singleModeSS}. For this, the modes of interest were driven by a sinusoidal signal locked by a phase-locked loop to the corresponding resonance peak. The resonator's response was measured for three different amplitudes per mode. Subsequently, the coefficient $b_{\textrm{F},i}$ was determined using the following relation,
\begin{equation}
	b_{\text{F},i} = \frac{A_{\text{out},i}Q_i}{A_{\text{in},i}C_m \Omega_i},
\end{equation}
which describes the resonator’s transfer function gain at resonance. Here, $A_{\text{in},i}$ and $A_{\text{out},i}$ represent the amplitudes of the sine waves applied to and measured from the resonator. $C_m$ describes the constant measurement amplification by the measurement apparatus, provided by the vendor.

\section{Observability and Controllability} \label{app:observe_control}
%Model descriptions in state space form offer a variety of methods to investigate the properties of the system. 
Two important system-theoretic properties of dynamic systems are observability and controllability. For completeness, we briefly explain both concepts but refer the reader to \cite{glasser1985control} for more details.

A dynamic system \eqref{eqn:LTI1} and \eqref{eqn:LTI2} is called fully \emph{observable} if its initial state $\mathbf{x}_0 = \mathbf{x}(0)$ can be retrieved from knowing the corresponding measurements $\mathbf{y}(t)$ for $0\leq t \leq T \le \infty$ \cite{kalman1960contributions}. For linear time-invariant systems, this is the case if and only if the observability matrix
\begin{equation}
    \mathcal{O}(\mathbf{A},\mathbf{C}) = \transpose{
        \left[
          \mathbf{C}   \quad
          \mathbf{C}\mathbf{A} \quad
          \mathbf{C}\mathbf{A}^2 \quad
          \cdots \quad
          \mathbf{C}\mathbf{A}^{n-1}
        \right]}
\end{equation}
has full rank, with $n$ being the number of states.
%By setting up the observability matrix $\mathbf{O}$ for the truncated NMR system from (\ref{eqn:MultharmSS}) one can ensure that the system is observable for either a pure displacement or a pure velocity measurement if the eigenfrequencys $\Omega_i$ of the different NMR modes are sufficiently far apart.

Similarly, a system \eqref{eqn:LTI1} is called fully \emph{controllable} if it can be steered from a state $\mathbf{x}(0) = \mathbf{0}$ to a desired state $\mathbf{x}(T) = \mathbf{x}_d$ in finite time by applying a suitable input $\mathbf{u}(t)$ for $0\leq t \leq T < \infty$. For linear time-invariant systems, this is the case if and only if the controllability matrix
\begin{equation}
    \mathcal{R}(\mathbf{A},\mathbf{B}) =
        \left[
          \mathbf{B}   \quad
          \mathbf{A}\mathbf{B} \quad
          \mathbf{A}^2\mathbf{B} \quad
          \cdots \quad
          \mathbf{A}^{n-1}\mathbf{B}
          \right]
\end{equation}
has full rank. 
Furthermore, a system is called \emph{stabilizable} if the subspaces associated with unstable eigenvalues of $\mathbf{A}$ is controllable. Equivalently, there exists a feedback law $\mathbf{u} = -\mathbf{K}_{\textrm{c}} \mathbf{x}$ such that the closed-loop dynamics $\mathbf{A} - \mathbf{B}\mathbf{K}_{\textrm{c}}$ is stable. Hence, a fully controllable system is also stabilizable.
%For an control input acting on the velocity state like an external force applied this can also be ensured for the NMR system.

\section{Controller implementation on FPGAs} \label{app:LQG_FPGA}
Optimal feedback control of the nanomechanical resonator requires a control concept (see section \ref{sec:opt_cntrl}) to run in real time. 
The equations (\ref{eqn:Klamfeqn}) and (\ref{eqn:lqrRaccati}) can be computed offline to obtain the steady-state observer and feedback gain $\mathbf{K}_{\textrm{f}}$ and $\mathbf{K}_{\textrm{c}}$. Hence, the resulting steady state LQG regulator equations follow as,
\begin{subequations}
	\begin{equation}
		\frac{\mathrm{d}}{\mathrm{d}t}\mathbf{x}_{\textrm{f}} = \underbrace{(\mathbf{A}-\mathbf{K}_{\textrm{f}}\mathbf{C} - \mathbf{B}\mathbf{K}_{c})}_{\mathbf{A}_{\textrm{f}}}\mathbf{x}_{\textrm{f}} + \mathbf{K}_{\textrm{f}}\mathbf{y},
	\end{equation}
	\begin{equation}
		\mathbf{u} = -\mathbf{K}_{c}\mathbf{x}_{\textrm{f}},
	\end{equation}
\end{subequations}
which have to be implemented on suitable real-time hardware such as FPGAs. This is done after temporal discretization,
% \section{Temporal Discretization of the LQG Regulator} \label{app:dt_LTI}
% For a linear time-invariant system of the form
% \begin{eqnarray*}
% 	\dot{\mathbf{x}}(t) = \mathbf{A}\mathbf{x}(t) + \mathbf{B}\mathbf{u}(t) \\
% 	\mathbf{y}(t) = \mathbf{C}\mathbf{x}(t) + \mathbf{D}\mathbf{u}(t)
% \end{eqnarray*}
% the general solution from some time $t_0$ to $t$ with external inputs can be expressed as
% \begin{equation*}
%     \mathbf{x}(t) = e^{\mathbf{A}(t-t_0)}\mathbf{x}(t_0) +\int_{t_0}^{t}e^{\mathbf{A}(t-\tau)}\mathbf{B}\mathbf{u}(\tau)\mathrm{d}\tau
% \end{equation*}
% with $=e^{\mathbf{A}t}$ being the matrix exponential of the dynamic matrix $\mathbf{A}$. Setting $t_0=kT_s$, $t=(k+1)T_s$ and $\mathbf{u}(\tau)=\mathbf{u}_k$ for $kT_s\leq \tau < (k+1)T_s$ for some sampling time $T_s$ and time index $k$ one gets
% \begin{equation*}
% 	\mathbf{x}(t) = e^{\mathbf{A}T_s}\mathbf{x}_k +\int_{kT_s}^{(k+1)T_s}e^{\mathbf{A}(k+1)T_s-\tau}\mathrm{d}\tau\mathbf{B}\mathbf{u}_k
% \end{equation*}
which results in the discretized state space system
\begin{eqnarray*}
        \mathbf{x}_{\textrm{f},k+1} &=& \mathbf{A}_{\textrm{df}}\mathbf{x}_{\textrm{f},k} + \mathbf{K}_{\textrm{df}}\mathbf{y}_{k},\\
       \mathbf{u}_{k} &=& -\mathbf{K}_{\mathbf{c}}\mathbf{x}_{\textrm{f},k}, 
\end{eqnarray*}
whereby the discrete matrices $\mathbf{A}_{\textrm{df}}$ and $\mathbf{K}_{\textrm{df}}$ are given by
\begin{eqnarray*}
    \mathbf{A}_{\textrm{df}} &=& e^{\mathbf{A}_{\textrm{f}}T_s}, \\
    \mathbf{K}_{\textrm{df}} &=& \mathbf{A}_{\textrm{f}}^{-1}(\mathbf{A}_{\textrm{df}}-\mathbf{I})\mathbf{K}_{\textrm{f}}
\end{eqnarray*}
The model is executed every 25 clock cycles at a clock frequency of \SI{125}{MHz}$=\frac{1}{T_s}$, resulting in an execution frequency of \SI{5}{MHz} as a trade-off between performance and numerical feasibility. For more details, the interested reader is referred to \cite{franklin1998digital}.

\section{Discrete-time Kalman-Bucy filter and RTS-Smoother} \label{app:dt_kf}
Since data is recorded as discrete samples in time, the post-processing procedure described in (\ref{chap:ImpEst}) is usually performed using a discrete-time version of the time-varying Kalman-Bucy filter. %The equations of the Kalman-Bucy filter can be discretized in time by integration of the corresponding continuous time matrix equations. 
The discrete time Kalman filter results in the set of difference equations
\begin{eqnarray*}
    \mathbf{x}_{\textrm{f},k+1} &=& \mathbf{A}_{\textrm{d}}\mathbf{x}_{\textrm{f},k} + \mathbf{B}_{\textrm{d}}\mathbf{u}_{k} + \mathbf{K}_{\textrm{f},k}(\mathbf{y}_k - \mathbf{C}\mathbf{x}_{\textrm{f},k})
\end{eqnarray*}
where the Kalman gain matrix is given by
\begin{equation*}
	\mathbf{K}_{\textrm{f},k} = \mathbf{A}_{\textrm{d}}\boldsymbol{\Sigma}_{\textrm{f},k}\transpose{\mathbf{C}}\left(\mathbf{C}\boldsymbol{\Sigma}_{\textrm{f},k}\transpose{\mathbf{C}} + \mathbf{R}_{\textrm{d}}\right)^{-1}
\end{equation*}
and the covariance update
\begin{equation*}
	\boldsymbol{\Sigma}_{\textrm{f},k+1} = \mathbf{A}_{\textrm{d}}\boldsymbol{\Sigma}_{\textrm{f},k}\transpose{\mathbf{A}_{\textrm{d}}} + \mathbf{Q}_{\textrm{d}} - 	\mathbf{K}_{\textrm{f},k}\left(\mathbf{C}\boldsymbol{\Sigma}_{\textrm{f},k}\transpose{\mathbf{A}_{\textrm{d}}}\right)
\end{equation*}
The discrete-time matrices $\mathbf{A}_{\textrm{d}}, \mathbf{B}_{\textrm{d}},\mathbf{Q}_{\textrm{d}}$, and $\mathbf{R}_{\textrm{d}}$ are given by \cite{franklin1998digital}
\begin{eqnarray*}
    \mathbf{A}_{\textrm{d}} &=& e^{\mathbf{A}T_s}, \\
    \mathbf{B}_{\textrm{d}} &=& \mathbf{A}^{-1}(\mathbf{A}_{\textrm{d}}-\mathbf{I})\mathbf{B},\\
	\mathbf{Q}_{\textrm{d}} &=& \int_{0}^{T_s}e^{\mathbf{A}\tau}\mathbf{G}\left(e^{ \mathbf{A} \tau}\mathbf{G} \right)^{\mathrm{T}} \mathrm{d}\tau,\\
	\mathbf{R}_{\textrm{d}} &=& \mathbf{R} / T_s,
\end{eqnarray*}
with $\mathbf{A}$, $\mathbf{B}$ and $\mathbf{G}$ from \eqref{eqn:LTI1}. 
%which can be computed by Van Loan's algorithm.

Analogous to the Kalman-Bucy filter above, the RTS-smoother differential equation can also be given in a discrete-time version as well.
The discrete-time RTS-smoother results in a set of difference equations applied backwards in time  \cite{lewis2017optimal} to obtain the smoothed state
\begin{equation*}
    \mathbf{x}_{\textrm{s},k} = \mathbf{x}_{\textrm{f},k} + \mathbf{G}_k\left(\mathbf{x}_{\textrm{s},k+1} - \mathbf{A}_{\textrm{d}}\mathbf{x}_{\textrm{f},k} - \mathbf{B}_{\textrm{d}}\mathbf{u}_{k}\right)
\end{equation*}
with smoother gain
\begin{equation*}
    \mathbf{G}_{k} = \mathbf{\Sigma}_{\textrm{f},k} \transpose{\mathbf{A}_{\textrm{d}}}\left( \mathbf{A}_{\textrm{d}}\boldsymbol{\Sigma}_{\textrm{f},k}\transpose{\mathbf{A}_{\textrm{d}}} + \mathbf{Q}_{\textrm{d}}\right)^{-1}
\end{equation*}
and the smoothed covariance
\begin{equation*}
    \mathbf{\Sigma}_{\textrm{s},k} = \mathbf{\Sigma}_{\textrm{f},k} + \mathbf{G}_k\left(\mathbf{\Sigma}_{\textrm{s},k+1} - \mathbf{A}_{\textrm{d}}\boldsymbol{\Sigma}_{\textrm{f},k}\transpose{\mathbf{A}_{\textrm{d}}} - \mathbf{Q}_{\textrm{d}}\right)\transpose{\mathbf{G}_{k}}.
\end{equation*}

% \section{From State Space to Transfer Function} 
% \label{app:tf_ss}
% A dynamic system in state space representation can always be uniquely be translated into an equivalent transfer function. 
% To transform a system of type

% into an input-output transfer function, the following transformation is used,
% \begin{equation*}
%     \mathbf{H}(s) = \frac{\mathbf{Y}(s)}{\mathbf{U}(s)} = \mathbf{C}(s\mathbf{I}-\mathbf{A})^{-1}\mathbf{B} + \mathbf{D}
% \end{equation*}
% with $\mathbf{I}$ being the identity matrix and $\mathbf{H}(s)$ being the transfer function from $\mathbf{U}(s)$ to $\mathbf{Y}(s)$. However, the reverse transformation from a transfer function into a state space model is not unique, since state space representations are not unique with respect to their input output behavior \cite{doyle2013feedback}.

% \section{Disturbance Model}
% \begin{figure}[h!]%{0.25\textwidth}
%     \begin{subfigure}{0.45\textwidth}
%         \centering
%         \caption{}
%         \input{figures/hit_statistics_without_noisepeak}
%         \label{fig:SpecCon}
%     \end{subfigure}
% \end{figure}

% The \nocite command causes all entries in a bibliography to be printed out
% whether or not they are actually referenced in the text. This is appropriate
% for the sample file to show the different styles of references, but authors
% most likely will not want to use it.
\nocite{*}

\bibliography{apssamp}% Produces the bibliography via BibTeX.

%apsrev4-2.bst 2019-01-14 (MD) hand-edited version of apsrev4-1.bst
%Control: key (0)
%Control: author (8) initials jnrlst
%Control: editor formatted (1) identically to author
%Control: production of article title (0) allowed
%Control: page (0) single
%Control: year (1) truncated
%Control: production of eprint (0) enabled
\begin{thebibliography}{64}%
\makeatletter
\providecommand \@ifxundefined [1]{%
 \@ifx{#1\undefined}
}%
\providecommand \@ifnum [1]{%
 \ifnum #1\expandafter \@firstoftwo
 \else \expandafter \@secondoftwo
 \fi
}%
\providecommand \@ifx [1]{%
 \ifx #1\expandafter \@firstoftwo
 \else \expandafter \@secondoftwo
 \fi
}%
\providecommand \natexlab [1]{#1}%
\providecommand \enquote  [1]{``#1''}%
\providecommand \bibnamefont  [1]{#1}%
\providecommand \bibfnamefont [1]{#1}%
\providecommand \citenamefont [1]{#1}%
\providecommand \href@noop [0]{\@secondoftwo}%
\providecommand \href [0]{\begingroup \@sanitize@url \@href}%
\providecommand \@href[1]{\@@startlink{#1}\@@href}%
\providecommand \@@href[1]{\endgroup#1\@@endlink}%
\providecommand \@sanitize@url [0]{\catcode `\\12\catcode `\$12\catcode `\&12\catcode `\#12\catcode `\^12\catcode `\_12\catcode `\%12\relax}%
\providecommand \@@startlink[1]{}%
\providecommand \@@endlink[0]{}%
\providecommand \url  [0]{\begingroup\@sanitize@url \@url }%
\providecommand \@url [1]{\endgroup\@href {#1}{\urlprefix }}%
\providecommand \urlprefix  [0]{URL }%
\providecommand \Eprint [0]{\href }%
\providecommand \doibase [0]{https://doi.org/}%
\providecommand \selectlanguage [0]{\@gobble}%
\providecommand \bibinfo  [0]{\@secondoftwo}%
\providecommand \bibfield  [0]{\@secondoftwo}%
\providecommand \translation [1]{[#1]}%
\providecommand \BibitemOpen [0]{}%
\providecommand \bibitemStop [0]{}%
\providecommand \bibitemNoStop [0]{.\EOS\space}%
\providecommand \EOS [0]{\spacefactor3000\relax}%
\providecommand \BibitemShut  [1]{\csname bibitem#1\endcsname}%
\let\auto@bib@innerbib\@empty
%</preamble>
\bibitem [{\citenamefont {Reinhardt}\ \emph {et~al.}(2016)\citenamefont {Reinhardt}, \citenamefont {M{\"u}ller}, \citenamefont {Bourassa},\ and\ \citenamefont {Sankey}}]{reinhardt2016ultralow}%
  \BibitemOpen
  \bibfield  {author} {\bibinfo {author} {\bibfnamefont {C.}~\bibnamefont {Reinhardt}}, \bibinfo {author} {\bibfnamefont {T.}~\bibnamefont {M{\"u}ller}}, \bibinfo {author} {\bibfnamefont {A.}~\bibnamefont {Bourassa}},\ and\ \bibinfo {author} {\bibfnamefont {J.~C.}\ \bibnamefont {Sankey}},\ }\bibfield  {title} {\bibinfo {title} {Ultralow-noise sin trampoline resonators for sensing and optomechanics},\ }\href@noop {} {\bibfield  {journal} {\bibinfo  {journal} {Physical Review X}\ }\textbf {\bibinfo {volume} {6}},\ \bibinfo {pages} {021001} (\bibinfo {year} {2016})}\BibitemShut {NoStop}%
\bibitem [{\citenamefont {Hanay}\ \emph {et~al.}(2012)\citenamefont {Hanay}, \citenamefont {Kelber}, \citenamefont {Naik}, \citenamefont {Chi}, \citenamefont {Hentz}, \citenamefont {Bullard}, \citenamefont {Colinet}, \citenamefont {Duraffourg},\ and\ \citenamefont {Roukes}}]{hanay2012single}%
  \BibitemOpen
  \bibfield  {author} {\bibinfo {author} {\bibfnamefont {M.~S.}\ \bibnamefont {Hanay}}, \bibinfo {author} {\bibfnamefont {S.}~\bibnamefont {Kelber}}, \bibinfo {author} {\bibfnamefont {A.}~\bibnamefont {Naik}}, \bibinfo {author} {\bibfnamefont {D.}~\bibnamefont {Chi}}, \bibinfo {author} {\bibfnamefont {S.}~\bibnamefont {Hentz}}, \bibinfo {author} {\bibfnamefont {E.}~\bibnamefont {Bullard}}, \bibinfo {author} {\bibfnamefont {E.}~\bibnamefont {Colinet}}, \bibinfo {author} {\bibfnamefont {L.}~\bibnamefont {Duraffourg}},\ and\ \bibinfo {author} {\bibfnamefont {M.}~\bibnamefont {Roukes}},\ }\bibfield  {title} {\bibinfo {title} {Single-protein nanomechanical mass spectrometry in real time},\ }\href@noop {} {\bibfield  {journal} {\bibinfo  {journal} {Nature nanotechnology}\ }\textbf {\bibinfo {volume} {7}},\ \bibinfo {pages} {602} (\bibinfo {year} {2012})}\BibitemShut {NoStop}%
\bibitem [{\citenamefont {Hanay}\ \emph {et~al.}(2015)\citenamefont {Hanay}, \citenamefont {Kelber}, \citenamefont {O'Connell}, \citenamefont {Mulvaney}, \citenamefont {Sader},\ and\ \citenamefont {Roukes}}]{hanay2015inertial}%
  \BibitemOpen
  \bibfield  {author} {\bibinfo {author} {\bibfnamefont {M.~S.}\ \bibnamefont {Hanay}}, \bibinfo {author} {\bibfnamefont {S.~I.}\ \bibnamefont {Kelber}}, \bibinfo {author} {\bibfnamefont {C.~D.}\ \bibnamefont {O'Connell}}, \bibinfo {author} {\bibfnamefont {P.}~\bibnamefont {Mulvaney}}, \bibinfo {author} {\bibfnamefont {J.~E.}\ \bibnamefont {Sader}},\ and\ \bibinfo {author} {\bibfnamefont {M.~L.}\ \bibnamefont {Roukes}},\ }\bibfield  {title} {\bibinfo {title} {Inertial imaging with nanomechanical systems},\ }\href@noop {} {\bibfield  {journal} {\bibinfo  {journal} {Nature nanotechnology}\ }\textbf {\bibinfo {volume} {10}},\ \bibinfo {pages} {339} (\bibinfo {year} {2015})}\BibitemShut {NoStop}%
\bibitem [{\citenamefont {Demir}(2021)}]{demir2021adaptive}%
  \BibitemOpen
  \bibfield  {author} {\bibinfo {author} {\bibfnamefont {A.}~\bibnamefont {Demir}},\ }\bibfield  {title} {\bibinfo {title} {Adaptive time-resolved mass spectrometry with nanomechanical resonant sensors},\ }\href@noop {} {\bibfield  {journal} {\bibinfo  {journal} {IEEE Sensors Journal}\ }\textbf {\bibinfo {volume} {21}},\ \bibinfo {pages} {27582} (\bibinfo {year} {2021})}\BibitemShut {NoStop}%
\bibitem [{\citenamefont {Li}\ \emph {et~al.}(2007)\citenamefont {Li}, \citenamefont {Tang},\ and\ \citenamefont {Roukes}}]{li2007ultra}%
  \BibitemOpen
  \bibfield  {author} {\bibinfo {author} {\bibfnamefont {M.}~\bibnamefont {Li}}, \bibinfo {author} {\bibfnamefont {H.~X.}\ \bibnamefont {Tang}},\ and\ \bibinfo {author} {\bibfnamefont {M.~L.}\ \bibnamefont {Roukes}},\ }\bibfield  {title} {\bibinfo {title} {Ultra-sensitive nems-based cantilevers for sensing, scanned probe and very high-frequency applications},\ }\href@noop {} {\bibfield  {journal} {\bibinfo  {journal} {Nature nanotechnology}\ }\textbf {\bibinfo {volume} {2}},\ \bibinfo {pages} {114} (\bibinfo {year} {2007})}\BibitemShut {NoStop}%
\bibitem [{\citenamefont {H{\"a}lg}\ \emph {et~al.}(2021)\citenamefont {H{\"a}lg}, \citenamefont {Gisler}, \citenamefont {Tsaturyan}, \citenamefont {Catalini}, \citenamefont {Grob}, \citenamefont {Krass}, \citenamefont {H{\'e}ritier}, \citenamefont {Mattiat}, \citenamefont {Thamm}, \citenamefont {Schirhagl} \emph {et~al.}}]{halg2021membrane}%
  \BibitemOpen
  \bibfield  {author} {\bibinfo {author} {\bibfnamefont {D.}~\bibnamefont {H{\"a}lg}}, \bibinfo {author} {\bibfnamefont {T.}~\bibnamefont {Gisler}}, \bibinfo {author} {\bibfnamefont {Y.}~\bibnamefont {Tsaturyan}}, \bibinfo {author} {\bibfnamefont {L.}~\bibnamefont {Catalini}}, \bibinfo {author} {\bibfnamefont {U.}~\bibnamefont {Grob}}, \bibinfo {author} {\bibfnamefont {M.-D.}\ \bibnamefont {Krass}}, \bibinfo {author} {\bibfnamefont {M.}~\bibnamefont {H{\'e}ritier}}, \bibinfo {author} {\bibfnamefont {H.}~\bibnamefont {Mattiat}}, \bibinfo {author} {\bibfnamefont {A.-K.}\ \bibnamefont {Thamm}}, \bibinfo {author} {\bibfnamefont {R.}~\bibnamefont {Schirhagl}}, \emph {et~al.},\ }\bibfield  {title} {\bibinfo {title} {Membrane-based scanning force microscopy},\ }\href@noop {} {\bibfield  {journal} {\bibinfo  {journal} {Physical Review Applied}\ }\textbf {\bibinfo {volume} {15}},\ \bibinfo {pages} {L021001} (\bibinfo {year} {2021})}\BibitemShut {NoStop}%
\bibitem [{\citenamefont {Gisler}\ \emph {et~al.}(2024)\citenamefont {Gisler}, \citenamefont {H\"alg}, \citenamefont {Dumont}, \citenamefont {Misra}, \citenamefont {Catalini}, \citenamefont {Langman}, \citenamefont {Schliesser}, \citenamefont {Degen},\ and\ \citenamefont {Eichler}}]{PhysRevApplied.22.044001}%
  \BibitemOpen
  \bibfield  {author} {\bibinfo {author} {\bibfnamefont {T.}~\bibnamefont {Gisler}}, \bibinfo {author} {\bibfnamefont {D.}~\bibnamefont {H\"alg}}, \bibinfo {author} {\bibfnamefont {V.}~\bibnamefont {Dumont}}, \bibinfo {author} {\bibfnamefont {S.}~\bibnamefont {Misra}}, \bibinfo {author} {\bibfnamefont {L.}~\bibnamefont {Catalini}}, \bibinfo {author} {\bibfnamefont {E.~C.}\ \bibnamefont {Langman}}, \bibinfo {author} {\bibfnamefont {A.}~\bibnamefont {Schliesser}}, \bibinfo {author} {\bibfnamefont {C.~L.}\ \bibnamefont {Degen}},\ and\ \bibinfo {author} {\bibfnamefont {A.}~\bibnamefont {Eichler}},\ }\bibfield  {title} {\bibinfo {title} {Enhancing membrane-based scanning force microscopy through an optical cavity},\ }\href {https://doi.org/10.1103/PhysRevApplied.22.044001} {\bibfield  {journal} {\bibinfo  {journal} {Phys. Rev. Appl.}\ }\textbf {\bibinfo {volume} {22}},\ \bibinfo {pages} {044001} (\bibinfo {year} {2024})}\BibitemShut {NoStop}%
\bibitem [{\citenamefont {Degen}\ \emph {et~al.}(2009)\citenamefont {Degen}, \citenamefont {Poggio}, \citenamefont {Mamin}, \citenamefont {Rettner},\ and\ \citenamefont {Rugar}}]{degen2009nanoscale}%
  \BibitemOpen
  \bibfield  {author} {\bibinfo {author} {\bibfnamefont {C.}~\bibnamefont {Degen}}, \bibinfo {author} {\bibfnamefont {M.}~\bibnamefont {Poggio}}, \bibinfo {author} {\bibfnamefont {H.}~\bibnamefont {Mamin}}, \bibinfo {author} {\bibfnamefont {C.}~\bibnamefont {Rettner}},\ and\ \bibinfo {author} {\bibfnamefont {D.}~\bibnamefont {Rugar}},\ }\bibfield  {title} {\bibinfo {title} {Nanoscale magnetic resonance imaging},\ }\href@noop {} {\bibfield  {journal} {\bibinfo  {journal} {Proceedings of the National Academy of Sciences}\ }\textbf {\bibinfo {volume} {106}},\ \bibinfo {pages} {1313} (\bibinfo {year} {2009})}\BibitemShut {NoStop}%
\bibitem [{\citenamefont {Grob}\ \emph {et~al.}(2019)\citenamefont {Grob}, \citenamefont {Krass}, \citenamefont {H{\'e}ritier}, \citenamefont {Pachlatko}, \citenamefont {Rhensius}, \citenamefont {Kosata}, \citenamefont {Moores}, \citenamefont {Takahashi}, \citenamefont {Eichler},\ and\ \citenamefont {Degen}}]{grob2019magnetic}%
  \BibitemOpen
  \bibfield  {author} {\bibinfo {author} {\bibfnamefont {U.}~\bibnamefont {Grob}}, \bibinfo {author} {\bibfnamefont {M.-D.}\ \bibnamefont {Krass}}, \bibinfo {author} {\bibfnamefont {M.}~\bibnamefont {H{\'e}ritier}}, \bibinfo {author} {\bibfnamefont {R.}~\bibnamefont {Pachlatko}}, \bibinfo {author} {\bibfnamefont {J.}~\bibnamefont {Rhensius}}, \bibinfo {author} {\bibfnamefont {J.}~\bibnamefont {Kosata}}, \bibinfo {author} {\bibfnamefont {B.}~\bibnamefont {Moores}}, \bibinfo {author} {\bibfnamefont {H.}~\bibnamefont {Takahashi}}, \bibinfo {author} {\bibfnamefont {A.}~\bibnamefont {Eichler}},\ and\ \bibinfo {author} {\bibfnamefont {C.~L.}\ \bibnamefont {Degen}},\ }\bibfield  {title} {\bibinfo {title} {Magnetic resonance force microscopy with a one-dimensional resolution of 0.9 nanometers},\ }\href@noop {} {\bibfield  {journal} {\bibinfo  {journal} {Nano letters}\ }\textbf {\bibinfo {volume} {19}},\ \bibinfo {pages} {7935} (\bibinfo {year} {2019})}\BibitemShut {NoStop}%
\bibitem [{\citenamefont {Bagci}\ \emph {et~al.}(2014)\citenamefont {Bagci}, \citenamefont {Simonsen}, \citenamefont {Schmid}, \citenamefont {Villanueva}, \citenamefont {Zeuthen}, \citenamefont {Appel}, \citenamefont {Taylor}, \citenamefont {S{\o}rensen}, \citenamefont {Usami}, \citenamefont {Schliesser} \emph {et~al.}}]{bagci2014optical}%
  \BibitemOpen
  \bibfield  {author} {\bibinfo {author} {\bibfnamefont {T.}~\bibnamefont {Bagci}}, \bibinfo {author} {\bibfnamefont {A.}~\bibnamefont {Simonsen}}, \bibinfo {author} {\bibfnamefont {S.}~\bibnamefont {Schmid}}, \bibinfo {author} {\bibfnamefont {L.~G.}\ \bibnamefont {Villanueva}}, \bibinfo {author} {\bibfnamefont {E.}~\bibnamefont {Zeuthen}}, \bibinfo {author} {\bibfnamefont {J.}~\bibnamefont {Appel}}, \bibinfo {author} {\bibfnamefont {J.~M.}\ \bibnamefont {Taylor}}, \bibinfo {author} {\bibfnamefont {A.}~\bibnamefont {S{\o}rensen}}, \bibinfo {author} {\bibfnamefont {K.}~\bibnamefont {Usami}}, \bibinfo {author} {\bibfnamefont {A.}~\bibnamefont {Schliesser}}, \emph {et~al.},\ }\bibfield  {title} {\bibinfo {title} {Optical detection of radio waves through a nanomechanical transducer},\ }\href@noop {} {\bibfield  {journal} {\bibinfo  {journal} {Nature}\ }\textbf {\bibinfo {volume} {507}},\ \bibinfo {pages} {81} (\bibinfo {year} {2014})}\BibitemShut {NoStop}%
\bibitem [{\citenamefont {Rossi}\ \emph {et~al.}(2018)\citenamefont {Rossi}, \citenamefont {Mason}, \citenamefont {Chen}, \citenamefont {Tsaturyan},\ and\ \citenamefont {Schliesser}}]{rossi2018measurement}%
  \BibitemOpen
  \bibfield  {author} {\bibinfo {author} {\bibfnamefont {M.}~\bibnamefont {Rossi}}, \bibinfo {author} {\bibfnamefont {D.}~\bibnamefont {Mason}}, \bibinfo {author} {\bibfnamefont {J.}~\bibnamefont {Chen}}, \bibinfo {author} {\bibfnamefont {Y.}~\bibnamefont {Tsaturyan}},\ and\ \bibinfo {author} {\bibfnamefont {A.}~\bibnamefont {Schliesser}},\ }\bibfield  {title} {\bibinfo {title} {Measurement-based quantum control of mechanical motion},\ }\href@noop {} {\bibfield  {journal} {\bibinfo  {journal} {Nature}\ }\textbf {\bibinfo {volume} {563}},\ \bibinfo {pages} {53} (\bibinfo {year} {2018})}\BibitemShut {NoStop}%
\bibitem [{\citenamefont {Deli{\'c}}\ \emph {et~al.}(2020)\citenamefont {Deli{\'c}}, \citenamefont {Reisenbauer}, \citenamefont {Dare}, \citenamefont {Grass}, \citenamefont {Vuleti{\'c}}, \citenamefont {Kiesel},\ and\ \citenamefont {Aspelmeyer}}]{delic2020cooling}%
  \BibitemOpen
  \bibfield  {author} {\bibinfo {author} {\bibfnamefont {U.}~\bibnamefont {Deli{\'c}}}, \bibinfo {author} {\bibfnamefont {M.}~\bibnamefont {Reisenbauer}}, \bibinfo {author} {\bibfnamefont {K.}~\bibnamefont {Dare}}, \bibinfo {author} {\bibfnamefont {D.}~\bibnamefont {Grass}}, \bibinfo {author} {\bibfnamefont {V.}~\bibnamefont {Vuleti{\'c}}}, \bibinfo {author} {\bibfnamefont {N.}~\bibnamefont {Kiesel}},\ and\ \bibinfo {author} {\bibfnamefont {M.}~\bibnamefont {Aspelmeyer}},\ }\bibfield  {title} {\bibinfo {title} {Cooling of a levitated nanoparticle to the motional quantum ground state},\ }\href@noop {} {\bibfield  {journal} {\bibinfo  {journal} {Science}\ }\textbf {\bibinfo {volume} {367}},\ \bibinfo {pages} {892} (\bibinfo {year} {2020})}\BibitemShut {NoStop}%
\bibitem [{\citenamefont {Seis}\ \emph {et~al.}(2022)\citenamefont {Seis}, \citenamefont {Capelle}, \citenamefont {Langman}, \citenamefont {Saarinen}, \citenamefont {Planz},\ and\ \citenamefont {Schliesser}}]{seis2022ground}%
  \BibitemOpen
  \bibfield  {author} {\bibinfo {author} {\bibfnamefont {Y.}~\bibnamefont {Seis}}, \bibinfo {author} {\bibfnamefont {T.}~\bibnamefont {Capelle}}, \bibinfo {author} {\bibfnamefont {E.}~\bibnamefont {Langman}}, \bibinfo {author} {\bibfnamefont {S.}~\bibnamefont {Saarinen}}, \bibinfo {author} {\bibfnamefont {E.}~\bibnamefont {Planz}},\ and\ \bibinfo {author} {\bibfnamefont {A.}~\bibnamefont {Schliesser}},\ }\bibfield  {title} {\bibinfo {title} {Ground state cooling of an ultracoherent electromechanical system},\ }\href@noop {} {\bibfield  {journal} {\bibinfo  {journal} {Nature communications}\ }\textbf {\bibinfo {volume} {13}},\ \bibinfo {pages} {1507} (\bibinfo {year} {2022})}\BibitemShut {NoStop}%
\bibitem [{\citenamefont {Huang}\ \emph {et~al.}(2024)\citenamefont {Huang}, \citenamefont {Beccari}, \citenamefont {Engelsen},\ and\ \citenamefont {Kippenberg}}]{huang2024room}%
  \BibitemOpen
  \bibfield  {author} {\bibinfo {author} {\bibfnamefont {G.}~\bibnamefont {Huang}}, \bibinfo {author} {\bibfnamefont {A.}~\bibnamefont {Beccari}}, \bibinfo {author} {\bibfnamefont {N.~J.}\ \bibnamefont {Engelsen}},\ and\ \bibinfo {author} {\bibfnamefont {T.~J.}\ \bibnamefont {Kippenberg}},\ }\bibfield  {title} {\bibinfo {title} {Room-temperature quantum optomechanics using an ultralow noise cavity},\ }\href@noop {} {\bibfield  {journal} {\bibinfo  {journal} {Nature}\ }\textbf {\bibinfo {volume} {626}},\ \bibinfo {pages} {512} (\bibinfo {year} {2024})}\BibitemShut {NoStop}%
\bibitem [{\citenamefont {Gonzfilez}\ and\ \citenamefont {Saulson}(1994)}]{gonzfilez1994brownian}%
  \BibitemOpen
  \bibfield  {author} {\bibinfo {author} {\bibfnamefont {G.~I.}\ \bibnamefont {Gonzfilez}}\ and\ \bibinfo {author} {\bibfnamefont {P.~R.}\ \bibnamefont {Saulson}},\ }\bibfield  {title} {\bibinfo {title} {Brownian motion of a mass suspended by an anelastic wire},\ }\href@noop {} {\bibfield  {journal} {\bibinfo  {journal} {Journal of the Acoustical Society of America}\ }\textbf {\bibinfo {volume} {96}},\ \bibinfo {pages} {207} (\bibinfo {year} {1994})}\BibitemShut {NoStop}%
\bibitem [{\citenamefont {Wang}\ \emph {et~al.}(2023)\citenamefont {Wang}, \citenamefont {Perez-Morelo}, \citenamefont {Ramer}, \citenamefont {Pavlidis}, \citenamefont {Schwartz}, \citenamefont {Yu}, \citenamefont {Ilic}, \citenamefont {Centrone},\ and\ \citenamefont {Aksyuk}}]{wang2023beating}%
  \BibitemOpen
  \bibfield  {author} {\bibinfo {author} {\bibfnamefont {M.}~\bibnamefont {Wang}}, \bibinfo {author} {\bibfnamefont {D.~J.}\ \bibnamefont {Perez-Morelo}}, \bibinfo {author} {\bibfnamefont {G.}~\bibnamefont {Ramer}}, \bibinfo {author} {\bibfnamefont {G.}~\bibnamefont {Pavlidis}}, \bibinfo {author} {\bibfnamefont {J.~J.}\ \bibnamefont {Schwartz}}, \bibinfo {author} {\bibfnamefont {L.}~\bibnamefont {Yu}}, \bibinfo {author} {\bibfnamefont {R.}~\bibnamefont {Ilic}}, \bibinfo {author} {\bibfnamefont {A.}~\bibnamefont {Centrone}},\ and\ \bibinfo {author} {\bibfnamefont {V.~A.}\ \bibnamefont {Aksyuk}},\ }\bibfield  {title} {\bibinfo {title} {Beating thermal noise in a dynamic signal measurement by a nanofabricated cavity optomechanical sensor},\ }\href@noop {} {\bibfield  {journal} {\bibinfo  {journal} {Science Advances}\ }\textbf {\bibinfo {volume} {9}},\ \bibinfo {pages} {eadf7595} (\bibinfo {year} {2023})}\BibitemShut {NoStop}%
\bibitem [{\citenamefont {Wieczorek}\ \emph {et~al.}(2015)\citenamefont {Wieczorek}, \citenamefont {Hofer}, \citenamefont {Hoelscher-Obermaier}, \citenamefont {Riedinger}, \citenamefont {Hammerer},\ and\ \citenamefont {Aspelmeyer}}]{wieczorek2015optimal}%
  \BibitemOpen
  \bibfield  {author} {\bibinfo {author} {\bibfnamefont {W.}~\bibnamefont {Wieczorek}}, \bibinfo {author} {\bibfnamefont {S.~G.}\ \bibnamefont {Hofer}}, \bibinfo {author} {\bibfnamefont {J.}~\bibnamefont {Hoelscher-Obermaier}}, \bibinfo {author} {\bibfnamefont {R.}~\bibnamefont {Riedinger}}, \bibinfo {author} {\bibfnamefont {K.}~\bibnamefont {Hammerer}},\ and\ \bibinfo {author} {\bibfnamefont {M.}~\bibnamefont {Aspelmeyer}},\ }\bibfield  {title} {\bibinfo {title} {Optimal state estimation for cavity optomechanical systems},\ }\href@noop {} {\bibfield  {journal} {\bibinfo  {journal} {Physical Review Letters}\ }\textbf {\bibinfo {volume} {114}},\ \bibinfo {pages} {223601} (\bibinfo {year} {2015})}\BibitemShut {NoStop}%
\bibitem [{\citenamefont {Kalman}\ and\ \citenamefont {Bucy}(1961)}]{kalmBu}%
  \BibitemOpen
  \bibfield  {author} {\bibinfo {author} {\bibfnamefont {R.~E.}\ \bibnamefont {Kalman}}\ and\ \bibinfo {author} {\bibfnamefont {R.~S.}\ \bibnamefont {Bucy}},\ }\bibfield  {title} {\bibinfo {title} {New results in linear filtering and prediction theory},\ }\href@noop {} {\bibfield  {journal} {\bibinfo  {journal} {Journal of Basic Engineering}\ }\textbf {\bibinfo {volume} {1}},\ \bibinfo {pages} {95} (\bibinfo {year} {1961})}\BibitemShut {NoStop}%
\bibitem [{\citenamefont {Rauch}\ \emph {et~al.}(1965)\citenamefont {Rauch}, \citenamefont {Tung},\ and\ \citenamefont {Striebel}}]{rtsSmo}%
  \BibitemOpen
  \bibfield  {author} {\bibinfo {author} {\bibfnamefont {H.~E.}\ \bibnamefont {Rauch}}, \bibinfo {author} {\bibfnamefont {F.}~\bibnamefont {Tung}},\ and\ \bibinfo {author} {\bibfnamefont {C.~T.}\ \bibnamefont {Striebel}},\ }\bibfield  {title} {\bibinfo {title} {Maximum likelihood estimates of linear dynamic systems},\ }\href@noop {} {\bibfield  {journal} {\bibinfo  {journal} {AIAA}\ }\textbf {\bibinfo {volume} {3}},\ \bibinfo {pages} {1445} (\bibinfo {year} {1965})}\BibitemShut {NoStop}%
\bibitem [{\citenamefont {S{\"a}rkk{\"a}}\ and\ \citenamefont {Svensson}(2023)}]{sarkka2023bayesian}%
  \BibitemOpen
  \bibfield  {author} {\bibinfo {author} {\bibfnamefont {S.}~\bibnamefont {S{\"a}rkk{\"a}}}\ and\ \bibinfo {author} {\bibfnamefont {L.}~\bibnamefont {Svensson}},\ }\href@noop {} {\emph {\bibinfo {title} {Bayesian filtering and smoothing}}},\ Vol.~\bibinfo {volume} {17}\ (\bibinfo  {publisher} {Cambridge university press},\ \bibinfo {year} {2023})\BibitemShut {NoStop}%
\bibitem [{\citenamefont {Schmid}\ \emph {et~al.}(2016)\citenamefont {Schmid}, \citenamefont {Villanueva},\ and\ \citenamefont {Roukes}}]{schmid2016fundamentals}%
  \BibitemOpen
  \bibfield  {author} {\bibinfo {author} {\bibfnamefont {S.}~\bibnamefont {Schmid}}, \bibinfo {author} {\bibfnamefont {L.~G.}\ \bibnamefont {Villanueva}},\ and\ \bibinfo {author} {\bibfnamefont {M.~L.}\ \bibnamefont {Roukes}},\ }\href@noop {} {\emph {\bibinfo {title} {Fundamentals of nanomechanical resonators}}},\ Vol.~\bibinfo {volume} {49}\ (\bibinfo  {publisher} {Springer},\ \bibinfo {year} {2016})\BibitemShut {NoStop}%
\bibitem [{\citenamefont {Erdogan}\ \emph {et~al.}(2024)\citenamefont {Erdogan}, \citenamefont {Baytekin}, \citenamefont {Coban},\ and\ \citenamefont {Demir}}]{Mete2024}%
  \BibitemOpen
  \bibfield  {author} {\bibinfo {author} {\bibfnamefont {M.}~\bibnamefont {Erdogan}}, \bibinfo {author} {\bibfnamefont {N.~B.}\ \bibnamefont {Baytekin}}, \bibinfo {author} {\bibfnamefont {S.~E.}\ \bibnamefont {Coban}},\ and\ \bibinfo {author} {\bibfnamefont {A.}~\bibnamefont {Demir}},\ }\bibfield  {title} {\bibinfo {title} {Machine learning and kalman filtering for nanomechanical mass spectrometry},\ }\href@noop {} {\bibfield  {journal} {\bibinfo  {journal} {IEEE Sensors Journal}\ }\textbf {\bibinfo {volume} {24}},\ \bibinfo {pages} {6303} (\bibinfo {year} {2024})}\BibitemShut {NoStop}%
\bibitem [{\citenamefont {Wang}\ \emph {et~al.}(2024)\citenamefont {Wang}, \citenamefont {Penny}, \citenamefont {Recoaro}, \citenamefont {Siegel}, \citenamefont {Tseng},\ and\ \citenamefont {Moore}}]{wang_mechanical_2024}%
  \BibitemOpen
  \bibfield  {author} {\bibinfo {author} {\bibfnamefont {J.}~\bibnamefont {Wang}}, \bibinfo {author} {\bibfnamefont {T.}~\bibnamefont {Penny}}, \bibinfo {author} {\bibfnamefont {J.}~\bibnamefont {Recoaro}}, \bibinfo {author} {\bibfnamefont {B.}~\bibnamefont {Siegel}}, \bibinfo {author} {\bibfnamefont {Y.-H.}\ \bibnamefont {Tseng}},\ and\ \bibinfo {author} {\bibfnamefont {D.~C.}\ \bibnamefont {Moore}},\ }\bibfield  {title} {\bibinfo {title} {Mechanical {Detection} of {Nuclear} {Decays}},\ }\href {https://doi.org/10.1103/PhysRevLett.133.023602} {\bibfield  {journal} {\bibinfo  {journal} {Physical Review Letters}\ }\textbf {\bibinfo {volume} {133}},\ \bibinfo {pages} {023602} (\bibinfo {year} {2024})}\BibitemShut {NoStop}%
\bibitem [{\citenamefont {Sansa}\ \emph {et~al.}(2020)\citenamefont {Sansa}, \citenamefont {Defoort}, \citenamefont {Brenac}, \citenamefont {Hermouet}, \citenamefont {Banniard}, \citenamefont {Fafin}, \citenamefont {Gely}, \citenamefont {Masselon}, \citenamefont {Favero}, \citenamefont {Jourdan} \emph {et~al.}}]{sansa2020optomechanical}%
  \BibitemOpen
  \bibfield  {author} {\bibinfo {author} {\bibfnamefont {M.}~\bibnamefont {Sansa}}, \bibinfo {author} {\bibfnamefont {M.}~\bibnamefont {Defoort}}, \bibinfo {author} {\bibfnamefont {A.}~\bibnamefont {Brenac}}, \bibinfo {author} {\bibfnamefont {M.}~\bibnamefont {Hermouet}}, \bibinfo {author} {\bibfnamefont {L.}~\bibnamefont {Banniard}}, \bibinfo {author} {\bibfnamefont {A.}~\bibnamefont {Fafin}}, \bibinfo {author} {\bibfnamefont {M.}~\bibnamefont {Gely}}, \bibinfo {author} {\bibfnamefont {C.}~\bibnamefont {Masselon}}, \bibinfo {author} {\bibfnamefont {I.}~\bibnamefont {Favero}}, \bibinfo {author} {\bibfnamefont {G.}~\bibnamefont {Jourdan}}, \emph {et~al.},\ }\bibfield  {title} {\bibinfo {title} {Optomechanical mass spectrometry},\ }\href@noop {} {\bibfield  {journal} {\bibinfo  {journal} {Nature Communications}\ }\textbf {\bibinfo {volume} {11}},\ \bibinfo {pages} {3781} (\bibinfo {year} {2020})}\BibitemShut {NoStop}%
\bibitem [{\citenamefont {Stassi}\ \emph {et~al.}(2019)\citenamefont {Stassi}, \citenamefont {De~Laurentis}, \citenamefont {Chakraborty}, \citenamefont {Bejtka}, \citenamefont {Chiodoni}, \citenamefont {Sader},\ and\ \citenamefont {Ricciardi}}]{stassi_large-scale_2019}%
  \BibitemOpen
  \bibfield  {author} {\bibinfo {author} {\bibfnamefont {S.}~\bibnamefont {Stassi}}, \bibinfo {author} {\bibfnamefont {G.}~\bibnamefont {De~Laurentis}}, \bibinfo {author} {\bibfnamefont {D.}~\bibnamefont {Chakraborty}}, \bibinfo {author} {\bibfnamefont {K.}~\bibnamefont {Bejtka}}, \bibinfo {author} {\bibfnamefont {A.}~\bibnamefont {Chiodoni}}, \bibinfo {author} {\bibfnamefont {J.~E.}\ \bibnamefont {Sader}},\ and\ \bibinfo {author} {\bibfnamefont {C.}~\bibnamefont {Ricciardi}},\ }\bibfield  {title} {\bibinfo {title} {Large-scale parallelization of nanomechanical mass spectrometry with weakly-coupled resonators},\ }\href {https://doi.org/10.1038/s41467-019-11647-2} {\bibfield  {journal} {\bibinfo  {journal} {Nature Communications}\ }\textbf {\bibinfo {volume} {10}},\ \bibinfo {pages} {3647} (\bibinfo {year} {2019})}\BibitemShut {NoStop}%
\bibitem [{\citenamefont {Ruz}\ \emph {et~al.}(2020)\citenamefont {Ruz}, \citenamefont {Malvar}, \citenamefont {Gil-Santos}, \citenamefont {Calleja},\ and\ \citenamefont {Tamayo}}]{ruz2020effect}%
  \BibitemOpen
  \bibfield  {author} {\bibinfo {author} {\bibfnamefont {J.}~\bibnamefont {Ruz}}, \bibinfo {author} {\bibfnamefont {O.}~\bibnamefont {Malvar}}, \bibinfo {author} {\bibfnamefont {E.}~\bibnamefont {Gil-Santos}}, \bibinfo {author} {\bibfnamefont {M.}~\bibnamefont {Calleja}},\ and\ \bibinfo {author} {\bibfnamefont {J.}~\bibnamefont {Tamayo}},\ }\bibfield  {title} {\bibinfo {title} {Effect of particle adsorption on the eigenfrequencies of nano-mechanical resonators},\ }\href@noop {} {\bibfield  {journal} {\bibinfo  {journal} {Journal of Applied Physics}\ }\textbf {\bibinfo {volume} {128}} (\bibinfo {year} {2020})}\BibitemShut {NoStop}%
\bibitem [{\citenamefont {Sage}\ \emph {et~al.}(2018)\citenamefont {Sage}, \citenamefont {Sansa}, \citenamefont {Fostner}, \citenamefont {Defoort}, \citenamefont {G{\'e}ly}, \citenamefont {Naik}, \citenamefont {Morel}, \citenamefont {Duraffourg}, \citenamefont {Roukes}, \citenamefont {Alava} \emph {et~al.}}]{sage2018single}%
  \BibitemOpen
  \bibfield  {author} {\bibinfo {author} {\bibfnamefont {E.}~\bibnamefont {Sage}}, \bibinfo {author} {\bibfnamefont {M.}~\bibnamefont {Sansa}}, \bibinfo {author} {\bibfnamefont {S.}~\bibnamefont {Fostner}}, \bibinfo {author} {\bibfnamefont {M.}~\bibnamefont {Defoort}}, \bibinfo {author} {\bibfnamefont {M.}~\bibnamefont {G{\'e}ly}}, \bibinfo {author} {\bibfnamefont {A.~K.}\ \bibnamefont {Naik}}, \bibinfo {author} {\bibfnamefont {R.}~\bibnamefont {Morel}}, \bibinfo {author} {\bibfnamefont {L.}~\bibnamefont {Duraffourg}}, \bibinfo {author} {\bibfnamefont {M.~L.}\ \bibnamefont {Roukes}}, \bibinfo {author} {\bibfnamefont {T.}~\bibnamefont {Alava}}, \emph {et~al.},\ }\bibfield  {title} {\bibinfo {title} {Single-particle mass spectrometry with arrays of frequency-addressed nanomechanical resonators},\ }\href@noop {} {\bibfield  {journal} {\bibinfo  {journal} {Nature communications}\ }\textbf {\bibinfo {volume} {9}},\ \bibinfo {pages} {3283} (\bibinfo {year} {2018})}\BibitemShut {NoStop}%
\bibitem [{\citenamefont {Naik}\ \emph {et~al.}(2009)\citenamefont {Naik}, \citenamefont {Hanay}, \citenamefont {Hiebert}, \citenamefont {Feng},\ and\ \citenamefont {Roukes}}]{naik2009towards}%
  \BibitemOpen
  \bibfield  {author} {\bibinfo {author} {\bibfnamefont {A.~K.}\ \bibnamefont {Naik}}, \bibinfo {author} {\bibfnamefont {M.}~\bibnamefont {Hanay}}, \bibinfo {author} {\bibfnamefont {W.}~\bibnamefont {Hiebert}}, \bibinfo {author} {\bibfnamefont {X.}~\bibnamefont {Feng}},\ and\ \bibinfo {author} {\bibfnamefont {M.~L.}\ \bibnamefont {Roukes}},\ }\bibfield  {title} {\bibinfo {title} {Towards single-molecule nanomechanical mass spectrometry},\ }\href@noop {} {\bibfield  {journal} {\bibinfo  {journal} {Nature nanotechnology}\ }\textbf {\bibinfo {volume} {4}},\ \bibinfo {pages} {445} (\bibinfo {year} {2009})}\BibitemShut {NoStop}%
\bibitem [{\citenamefont {Be{\v{s}}i{\'c}}\ \emph {et~al.}(2024)\citenamefont {Be{\v{s}}i{\'c}}, \citenamefont {Deutschmann-Olek}, \citenamefont {Me{\v{s}}i{\'c}}, \citenamefont {Kanellopulos},\ and\ \citenamefont {Schmid}}]{bevsic2024optimized}%
  \BibitemOpen
  \bibfield  {author} {\bibinfo {author} {\bibfnamefont {H.}~\bibnamefont {Be{\v{s}}i{\'c}}}, \bibinfo {author} {\bibfnamefont {A.}~\bibnamefont {Deutschmann-Olek}}, \bibinfo {author} {\bibfnamefont {K.}~\bibnamefont {Me{\v{s}}i{\'c}}}, \bibinfo {author} {\bibfnamefont {K.}~\bibnamefont {Kanellopulos}},\ and\ \bibinfo {author} {\bibfnamefont {S.}~\bibnamefont {Schmid}},\ }\bibfield  {title} {\bibinfo {title} {Optimized signal estimation in nanomechanical photothermal sensing via thermal response modelling and kalman filtering},\ }\href@noop {} {\bibfield  {journal} {\bibinfo  {journal} {arXiv preprint arXiv:2405.15938}\ } (\bibinfo {year} {2024})}\BibitemShut {NoStop}%
\bibitem [{\citenamefont {Magrini}\ \emph {et~al.}(2021)\citenamefont {Magrini}, \citenamefont {Rosenzweig}, \citenamefont {Bach}, \citenamefont {Deutschmann-Olek}, \citenamefont {Hofer}, \citenamefont {Hong}, \citenamefont {Kiesel}, \citenamefont {Kugi},\ and\ \citenamefont {Aspelmeyer}}]{magrini2021real}%
  \BibitemOpen
  \bibfield  {author} {\bibinfo {author} {\bibfnamefont {L.}~\bibnamefont {Magrini}}, \bibinfo {author} {\bibfnamefont {P.}~\bibnamefont {Rosenzweig}}, \bibinfo {author} {\bibfnamefont {C.}~\bibnamefont {Bach}}, \bibinfo {author} {\bibfnamefont {A.}~\bibnamefont {Deutschmann-Olek}}, \bibinfo {author} {\bibfnamefont {S.~G.}\ \bibnamefont {Hofer}}, \bibinfo {author} {\bibfnamefont {S.}~\bibnamefont {Hong}}, \bibinfo {author} {\bibfnamefont {N.}~\bibnamefont {Kiesel}}, \bibinfo {author} {\bibfnamefont {A.}~\bibnamefont {Kugi}},\ and\ \bibinfo {author} {\bibfnamefont {M.}~\bibnamefont {Aspelmeyer}},\ }\bibfield  {title} {\bibinfo {title} {Real-time optimal quantum control of mechanical motion at room temperature},\ }\href@noop {} {\bibfield  {journal} {\bibinfo  {journal} {Nature}\ }\textbf {\bibinfo {volume} {595}},\ \bibinfo {pages} {373} (\bibinfo {year} {2021})}\BibitemShut {NoStop}%
\bibitem [{\citenamefont {Barker}\ \emph {et~al.}(2024)\citenamefont {Barker}, \citenamefont {Carney}, \citenamefont {LeBrun}, \citenamefont {Moore},\ and\ \citenamefont {Taylor}}]{barker2024collision}%
  \BibitemOpen
  \bibfield  {author} {\bibinfo {author} {\bibfnamefont {D.~S.}\ \bibnamefont {Barker}}, \bibinfo {author} {\bibfnamefont {D.}~\bibnamefont {Carney}}, \bibinfo {author} {\bibfnamefont {T.~W.}\ \bibnamefont {LeBrun}}, \bibinfo {author} {\bibfnamefont {D.~C.}\ \bibnamefont {Moore}},\ and\ \bibinfo {author} {\bibfnamefont {J.~M.}\ \bibnamefont {Taylor}},\ }\bibfield  {title} {\bibinfo {title} {Collision-resolved pressure sensing},\ }\href@noop {} {\bibfield  {journal} {\bibinfo  {journal} {Physical Review A}\ }\textbf {\bibinfo {volume} {109}},\ \bibinfo {pages} {042616} (\bibinfo {year} {2024})}\BibitemShut {NoStop}%
\bibitem [{\citenamefont {Carney}\ \emph {et~al.}(2021)\citenamefont {Carney}, \citenamefont {Krnjaic}, \citenamefont {Moore}, \citenamefont {Regal} \emph {et~al.}}]{carney_mechanical_2021}%
  \BibitemOpen
  \bibfield  {author} {\bibinfo {author} {\bibfnamefont {D.}~\bibnamefont {Carney}}, \bibinfo {author} {\bibfnamefont {G.}~\bibnamefont {Krnjaic}}, \bibinfo {author} {\bibfnamefont {D.~C.}\ \bibnamefont {Moore}}, \bibinfo {author} {\bibfnamefont {C.~A.}\ \bibnamefont {Regal}}, \emph {et~al.},\ }\bibfield  {title} {\bibinfo {title} {Mechanical quantum sensing in the search for dark matter},\ }\href {https://doi.org/10.1088/2058-9565/abcfcd} {\bibfield  {journal} {\bibinfo  {journal} {Quantum Science and Technology}\ }\textbf {\bibinfo {volume} {6}},\ \bibinfo {pages} {024002} (\bibinfo {year} {2021})}\BibitemShut {NoStop}%
\bibitem [{\citenamefont {Li}\ \emph {et~al.}(2023)\citenamefont {Li}, \citenamefont {Cai}, \citenamefont {Hao}, \citenamefont {Smith},\ and\ \citenamefont {Jiang}}]{li_online_2023}%
  \BibitemOpen
  \bibfield  {author} {\bibinfo {author} {\bibfnamefont {X.}~\bibnamefont {Li}}, \bibinfo {author} {\bibfnamefont {R.}~\bibnamefont {Cai}}, \bibinfo {author} {\bibfnamefont {J.}~\bibnamefont {Hao}}, \bibinfo {author} {\bibfnamefont {J.~N.}\ \bibnamefont {Smith}},\ and\ \bibinfo {author} {\bibfnamefont {J.}~\bibnamefont {Jiang}},\ }\bibfield  {title} {\bibinfo {title} {Online detection of airborne nanoparticle composition with mass spectrometry: {Recent} advances, challenges, and opportunities},\ }\href {https://doi.org/10.1016/j.trac.2023.117195} {\bibfield  {journal} {\bibinfo  {journal} {TrAC Trends in Analytical Chemistry}\ }\textbf {\bibinfo {volume} {166}},\ \bibinfo {pages} {117195} (\bibinfo {year} {2023})}\BibitemShut {NoStop}%
\bibitem [{\citenamefont {Bennett}\ \emph {et~al.}(2023)\citenamefont {Bennett}, \citenamefont {Stephenson}, \citenamefont {Rose},\ and\ \citenamefont {Darmanis}}]{bennett_single-cell_2023}%
  \BibitemOpen
  \bibfield  {author} {\bibinfo {author} {\bibfnamefont {H.~M.}\ \bibnamefont {Bennett}}, \bibinfo {author} {\bibfnamefont {W.}~\bibnamefont {Stephenson}}, \bibinfo {author} {\bibfnamefont {C.~M.}\ \bibnamefont {Rose}},\ and\ \bibinfo {author} {\bibfnamefont {S.}~\bibnamefont {Darmanis}},\ }\bibfield  {title} {\bibinfo {title} {Single-cell proteomics enabled by next-generation sequencing or mass spectrometry},\ }\href {https://doi.org/10.1038/s41592-023-01791-5} {\bibfield  {journal} {\bibinfo  {journal} {Nature Methods}\ }\textbf {\bibinfo {volume} {20}},\ \bibinfo {pages} {363} (\bibinfo {year} {2023})}\BibitemShut {NoStop}%
\bibitem [{\citenamefont {Hopkins}\ \emph {et~al.}(2003)\citenamefont {Hopkins}, \citenamefont {Jacobs}, \citenamefont {Habib},\ and\ \citenamefont {Schwab}}]{hopkins2003feedback}%
  \BibitemOpen
  \bibfield  {author} {\bibinfo {author} {\bibfnamefont {A.}~\bibnamefont {Hopkins}}, \bibinfo {author} {\bibfnamefont {K.}~\bibnamefont {Jacobs}}, \bibinfo {author} {\bibfnamefont {S.}~\bibnamefont {Habib}},\ and\ \bibinfo {author} {\bibfnamefont {K.}~\bibnamefont {Schwab}},\ }\bibfield  {title} {\bibinfo {title} {Feedback cooling of a nanomechanical resonator},\ }\href@noop {} {\bibfield  {journal} {\bibinfo  {journal} {Physical Review B}\ }\textbf {\bibinfo {volume} {68}},\ \bibinfo {pages} {235328} (\bibinfo {year} {2003})}\BibitemShut {NoStop}%
\bibitem [{\citenamefont {Kleckner}\ and\ \citenamefont {Bouwmeester}(2006)}]{kleckner2006sub}%
  \BibitemOpen
  \bibfield  {author} {\bibinfo {author} {\bibfnamefont {D.}~\bibnamefont {Kleckner}}\ and\ \bibinfo {author} {\bibfnamefont {D.}~\bibnamefont {Bouwmeester}},\ }\bibfield  {title} {\bibinfo {title} {Sub-kelvin optical cooling of a micromechanical resonator},\ }\href@noop {} {\bibfield  {journal} {\bibinfo  {journal} {Nature}\ }\textbf {\bibinfo {volume} {444}},\ \bibinfo {pages} {75} (\bibinfo {year} {2006})}\BibitemShut {NoStop}%
\bibitem [{\citenamefont {Zhang}\ and\ \citenamefont {Mølmer}(2017)}]{zhang_prediction_2017}%
  \BibitemOpen
  \bibfield  {author} {\bibinfo {author} {\bibfnamefont {J.}~\bibnamefont {Zhang}}\ and\ \bibinfo {author} {\bibfnamefont {K.}~\bibnamefont {Mølmer}},\ }\bibfield  {title} {\bibinfo {title} {Prediction and retrodiction with continuously monitored {Gaussian} states},\ }\href {https://doi.org/10.1103/PhysRevA.96.062131} {\bibfield  {journal} {\bibinfo  {journal} {Physical Review A}\ }\textbf {\bibinfo {volume} {96}},\ \bibinfo {pages} {062131} (\bibinfo {year} {2017})}\BibitemShut {NoStop}%
\bibitem [{\citenamefont {Bao}\ \emph {et~al.}(2020)\citenamefont {Bao}, \citenamefont {Jin}, \citenamefont {Duan}, \citenamefont {Jia}, \citenamefont {Mølmer}, \citenamefont {Shen},\ and\ \citenamefont {Xiao}}]{bao_retrodiction_2020}%
  \BibitemOpen
  \bibfield  {author} {\bibinfo {author} {\bibfnamefont {H.}~\bibnamefont {Bao}}, \bibinfo {author} {\bibfnamefont {S.}~\bibnamefont {Jin}}, \bibinfo {author} {\bibfnamefont {J.}~\bibnamefont {Duan}}, \bibinfo {author} {\bibfnamefont {S.}~\bibnamefont {Jia}}, \bibinfo {author} {\bibfnamefont {K.}~\bibnamefont {Mølmer}}, \bibinfo {author} {\bibfnamefont {H.}~\bibnamefont {Shen}},\ and\ \bibinfo {author} {\bibfnamefont {Y.}~\bibnamefont {Xiao}},\ }\bibfield  {title} {\bibinfo {title} {Retrodiction beyond the {Heisenberg} uncertainty relation},\ }\href {https://doi.org/10.1038/s41467-020-19495-1} {\bibfield  {journal} {\bibinfo  {journal} {Nature Communications}\ }\textbf {\bibinfo {volume} {11}},\ \bibinfo {pages} {5658} (\bibinfo {year} {2020})}\BibitemShut {NoStop}%
\bibitem [{\citenamefont {Lammers}\ and\ \citenamefont {Hammerer}(2024)}]{lammers_quantum_2024}%
  \BibitemOpen
  \bibfield  {author} {\bibinfo {author} {\bibfnamefont {J.}~\bibnamefont {Lammers}}\ and\ \bibinfo {author} {\bibfnamefont {K.}~\bibnamefont {Hammerer}},\ }\bibfield  {title} {\bibinfo {title} {Quantum retrodiction in {Gaussian} systems and applications in optomechanics},\ }\bibfield  {journal} {\bibinfo  {journal} {Frontiers in Quantum Science and Technology}\ }\textbf {\bibinfo {volume} {2}},\ \href {https://doi.org/10.3389/frqst.2023.1294905} {10.3389/frqst.2023.1294905} (\bibinfo {year} {2024})\BibitemShut {NoStop}%
\bibitem [{\citenamefont {Badawi}\ \emph {et~al.}(1979)\citenamefont {Badawi}, \citenamefont {Lindquist},\ and\ \citenamefont {Pavon}}]{badawi1979stochastic}%
  \BibitemOpen
  \bibfield  {author} {\bibinfo {author} {\bibfnamefont {F.}~\bibnamefont {Badawi}}, \bibinfo {author} {\bibfnamefont {A.}~\bibnamefont {Lindquist}},\ and\ \bibinfo {author} {\bibfnamefont {M.}~\bibnamefont {Pavon}},\ }\bibfield  {title} {\bibinfo {title} {A stochastic realization approach to the smoothing problem},\ }\href@noop {} {\bibfield  {journal} {\bibinfo  {journal} {IEEE Transactions on Automatic Control}\ }\textbf {\bibinfo {volume} {24}},\ \bibinfo {pages} {878} (\bibinfo {year} {1979})}\BibitemShut {NoStop}%
\bibitem [{\citenamefont {Kubo}(1966)}]{kubo1966fluctuation}%
  \BibitemOpen
  \bibfield  {author} {\bibinfo {author} {\bibfnamefont {R.}~\bibnamefont {Kubo}},\ }\bibfield  {title} {\bibinfo {title} {The fluctuation-dissipation theorem},\ }\href@noop {} {\bibfield  {journal} {\bibinfo  {journal} {Reports on Progress in Physics}\ }\textbf {\bibinfo {volume} {29}},\ \bibinfo {pages} {255} (\bibinfo {year} {1966})}\BibitemShut {NoStop}%
\bibitem [{\citenamefont {Callen}\ and\ \citenamefont {Welton}(1951)}]{callen1951irreversibility}%
  \BibitemOpen
  \bibfield  {author} {\bibinfo {author} {\bibfnamefont {H.~B.}\ \bibnamefont {Callen}}\ and\ \bibinfo {author} {\bibfnamefont {T.~A.}\ \bibnamefont {Welton}},\ }\bibfield  {title} {\bibinfo {title} {Irreversibility and generalized noise},\ }\href@noop {} {\bibfield  {journal} {\bibinfo  {journal} {Physical Review}\ }\textbf {\bibinfo {volume} {83}},\ \bibinfo {pages} {34} (\bibinfo {year} {1951})}\BibitemShut {NoStop}%
\bibitem [{\citenamefont {Luenberger}(1964)}]{luenberger1964observing}%
  \BibitemOpen
  \bibfield  {author} {\bibinfo {author} {\bibfnamefont {D.~G.}\ \bibnamefont {Luenberger}},\ }\bibfield  {title} {\bibinfo {title} {Observing the state of a linear system},\ }\href@noop {} {\bibfield  {journal} {\bibinfo  {journal} {IEEE Transactions on Military Electronics}\ }\textbf {\bibinfo {volume} {8}},\ \bibinfo {pages} {74} (\bibinfo {year} {1964})}\BibitemShut {NoStop}%
\bibitem [{\citenamefont {Fraser}\ and\ \citenamefont {Potter}(1969)}]{fraser1969optimum}%
  \BibitemOpen
  \bibfield  {author} {\bibinfo {author} {\bibfnamefont {D.}~\bibnamefont {Fraser}}\ and\ \bibinfo {author} {\bibfnamefont {J.}~\bibnamefont {Potter}},\ }\bibfield  {title} {\bibinfo {title} {The optimum linear smoother as a combination of two optimum linear filters},\ }\href@noop {} {\bibfield  {journal} {\bibinfo  {journal} {IEEE Transactions on Automatic Control}\ }\textbf {\bibinfo {volume} {14}},\ \bibinfo {pages} {387} (\bibinfo {year} {1969})}\BibitemShut {NoStop}%
\bibitem [{\citenamefont {Mayne}(1966)}]{mayne1966solution}%
  \BibitemOpen
  \bibfield  {author} {\bibinfo {author} {\bibfnamefont {D.~Q.}\ \bibnamefont {Mayne}},\ }\bibfield  {title} {\bibinfo {title} {A solution of the smoothing problem for linear dynamic systems},\ }\href@noop {} {\bibfield  {journal} {\bibinfo  {journal} {Automatica}\ }\textbf {\bibinfo {volume} {4}},\ \bibinfo {pages} {73} (\bibinfo {year} {1966})}\BibitemShut {NoStop}%
\bibitem [{\citenamefont {Athans}(1971)}]{athans1971role}%
  \BibitemOpen
  \bibfield  {author} {\bibinfo {author} {\bibfnamefont {M.}~\bibnamefont {Athans}},\ }\bibfield  {title} {\bibinfo {title} {The role and use of the stochastic linear-quadratic-gaussian problem in control system design},\ }\href@noop {} {\bibfield  {journal} {\bibinfo  {journal} {IEEE Transactions on Automatic Control}\ }\textbf {\bibinfo {volume} {16}},\ \bibinfo {pages} {529} (\bibinfo {year} {1971})}\BibitemShut {NoStop}%
\bibitem [{\citenamefont {Bode}\ and\ \citenamefont {Shannon}(1950)}]{bode1950simplified}%
  \BibitemOpen
  \bibfield  {author} {\bibinfo {author} {\bibfnamefont {H.~W.}\ \bibnamefont {Bode}}\ and\ \bibinfo {author} {\bibfnamefont {C.~E.}\ \bibnamefont {Shannon}},\ }\bibfield  {title} {\bibinfo {title} {A simplified derivation of linear least square smoothing and prediction theory},\ }\href@noop {} {\bibfield  {journal} {\bibinfo  {journal} {Proceedings of the IRE}\ }\textbf {\bibinfo {volume} {38}},\ \bibinfo {pages} {417} (\bibinfo {year} {1950})}\BibitemShut {NoStop}%
\bibitem [{\citenamefont {Tebbenjohanns}\ \emph {et~al.}(2021)\citenamefont {Tebbenjohanns}, \citenamefont {Mattana}, \citenamefont {Rossi}, \citenamefont {Frimmer},\ and\ \citenamefont {Novotny}}]{tebbenjohanns_quantum_2021}%
  \BibitemOpen
  \bibfield  {author} {\bibinfo {author} {\bibfnamefont {F.}~\bibnamefont {Tebbenjohanns}}, \bibinfo {author} {\bibfnamefont {M.~L.}\ \bibnamefont {Mattana}}, \bibinfo {author} {\bibfnamefont {M.}~\bibnamefont {Rossi}}, \bibinfo {author} {\bibfnamefont {M.}~\bibnamefont {Frimmer}},\ and\ \bibinfo {author} {\bibfnamefont {L.}~\bibnamefont {Novotny}},\ }\bibfield  {title} {\bibinfo {title} {Quantum control of a nanoparticle optically levitated in cryogenic free space},\ }\href {https://doi.org/10.1038/s41586-021-03617-w} {\bibfield  {journal} {\bibinfo  {journal} {Nature}\ }\textbf {\bibinfo {volume} {595}},\ \bibinfo {pages} {378} (\bibinfo {year} {2021})}\BibitemShut {NoStop}%
\bibitem [{\citenamefont {Xia}\ \emph {et~al.}(2024)\citenamefont {Xia}, \citenamefont {Huang}, \citenamefont {Beccari}, \citenamefont {Zicoschi}, \citenamefont {Arabmoheghi}, \citenamefont {Engelsen},\ and\ \citenamefont {Kippenberg}}]{xia_motional_2024}%
  \BibitemOpen
  \bibfield  {author} {\bibinfo {author} {\bibfnamefont {Y.}~\bibnamefont {Xia}}, \bibinfo {author} {\bibfnamefont {G.}~\bibnamefont {Huang}}, \bibinfo {author} {\bibfnamefont {A.}~\bibnamefont {Beccari}}, \bibinfo {author} {\bibfnamefont {A.}~\bibnamefont {Zicoschi}}, \bibinfo {author} {\bibfnamefont {A.}~\bibnamefont {Arabmoheghi}}, \bibinfo {author} {\bibfnamefont {N.~J.}\ \bibnamefont {Engelsen}},\ and\ \bibinfo {author} {\bibfnamefont {T.~J.}\ \bibnamefont {Kippenberg}},\ }\href {https://doi.org/10.48550/arXiv.2408.06498} {\bibinfo {title} {Motional sideband asymmetry of a solid-state mechanical resonator at room temperature}} (\bibinfo {year} {2024})\BibitemShut {NoStop}%
\bibitem [{\citenamefont {Tsaturyan}\ \emph {et~al.}(2017)\citenamefont {Tsaturyan}, \citenamefont {Barg}, \citenamefont {Polzik},\ and\ \citenamefont {Schliesser}}]{tsaturyan2017ultracoherent}%
  \BibitemOpen
  \bibfield  {author} {\bibinfo {author} {\bibfnamefont {Y.}~\bibnamefont {Tsaturyan}}, \bibinfo {author} {\bibfnamefont {A.}~\bibnamefont {Barg}}, \bibinfo {author} {\bibfnamefont {E.~S.}\ \bibnamefont {Polzik}},\ and\ \bibinfo {author} {\bibfnamefont {A.}~\bibnamefont {Schliesser}},\ }\bibfield  {title} {\bibinfo {title} {Ultracoherent nanomechanical resonators via soft clamping and dissipation dilution},\ }\href@noop {} {\bibfield  {journal} {\bibinfo  {journal} {Nature Nanotechnology}\ }\textbf {\bibinfo {volume} {12}},\ \bibinfo {pages} {776} (\bibinfo {year} {2017})}\BibitemShut {NoStop}%
\bibitem [{\citenamefont {Engelsen}\ \emph {et~al.}(2024)\citenamefont {Engelsen}, \citenamefont {Beccari},\ and\ \citenamefont {Kippenberg}}]{engelsen2024ultrahigh}%
  \BibitemOpen
  \bibfield  {author} {\bibinfo {author} {\bibfnamefont {N.~J.}\ \bibnamefont {Engelsen}}, \bibinfo {author} {\bibfnamefont {A.}~\bibnamefont {Beccari}},\ and\ \bibinfo {author} {\bibfnamefont {T.~J.}\ \bibnamefont {Kippenberg}},\ }\bibfield  {title} {\bibinfo {title} {Ultrahigh-quality-factor micro-and nanomechanical resonators using dissipation dilution},\ }\href@noop {} {\bibfield  {journal} {\bibinfo  {journal} {Nature Nanotechnology}\ ,\ \bibinfo {pages} {1}} (\bibinfo {year} {2024})}\BibitemShut {NoStop}%
\bibitem [{\citenamefont {Yang}\ \emph {et~al.}(2006)\citenamefont {Yang}, \citenamefont {Callegari}, \citenamefont {Feng}, \citenamefont {Ekinci},\ and\ \citenamefont {Roukes}}]{yang_zeptogram-scale_2006}%
  \BibitemOpen
  \bibfield  {author} {\bibinfo {author} {\bibfnamefont {Y.~T.}\ \bibnamefont {Yang}}, \bibinfo {author} {\bibfnamefont {C.}~\bibnamefont {Callegari}}, \bibinfo {author} {\bibfnamefont {X.~L.}\ \bibnamefont {Feng}}, \bibinfo {author} {\bibfnamefont {K.~L.}\ \bibnamefont {Ekinci}},\ and\ \bibinfo {author} {\bibfnamefont {M.~L.}\ \bibnamefont {Roukes}},\ }\bibfield  {title} {\bibinfo {title} {Zeptogram-{Scale} {Nanomechanical} {Mass} {Sensing}},\ }\href {https://doi.org/10.1021/nl052134m} {\bibfield  {journal} {\bibinfo  {journal} {Nano Letters}\ }\textbf {\bibinfo {volume} {6}},\ \bibinfo {pages} {583} (\bibinfo {year} {2006})}\BibitemShut {NoStop}%
\bibitem [{\citenamefont {Hauer}\ \emph {et~al.}(2013)\citenamefont {Hauer}, \citenamefont {Doolin}, \citenamefont {Beach},\ and\ \citenamefont {Davis}}]{hauer2013general}%
  \BibitemOpen
  \bibfield  {author} {\bibinfo {author} {\bibfnamefont {B.}~\bibnamefont {Hauer}}, \bibinfo {author} {\bibfnamefont {C.}~\bibnamefont {Doolin}}, \bibinfo {author} {\bibfnamefont {K.}~\bibnamefont {Beach}},\ and\ \bibinfo {author} {\bibfnamefont {J.}~\bibnamefont {Davis}},\ }\bibfield  {title} {\bibinfo {title} {A general procedure for thermomechanical calibration of nano/micro-mechanical resonators},\ }\href@noop {} {\bibfield  {journal} {\bibinfo  {journal} {Annals of Physics}\ }\textbf {\bibinfo {volume} {339}},\ \bibinfo {pages} {181} (\bibinfo {year} {2013})}\BibitemShut {NoStop}%
\bibitem [{\citenamefont {Glasser}(1985)}]{glasser1985control}%
  \BibitemOpen
  \bibfield  {author} {\bibinfo {author} {\bibfnamefont {W.}~\bibnamefont {Glasser}},\ }\href@noop {} {\emph {\bibinfo {title} {Control theory}}}\ (\bibinfo  {publisher} {Harper and Row New York},\ \bibinfo {year} {1985})\BibitemShut {NoStop}%
\bibitem [{\citenamefont {Kalman}\ \emph {et~al.}(1960)\citenamefont {Kalman} \emph {et~al.}}]{kalman1960contributions}%
  \BibitemOpen
  \bibfield  {author} {\bibinfo {author} {\bibfnamefont {R.~E.}\ \bibnamefont {Kalman}} \emph {et~al.},\ }\bibfield  {title} {\bibinfo {title} {Contributions to the theory of optimal control},\ }\href@noop {} {\bibfield  {journal} {\bibinfo  {journal} {Bol. soc. mat. mexicana}\ }\textbf {\bibinfo {volume} {5}},\ \bibinfo {pages} {102} (\bibinfo {year} {1960})}\BibitemShut {NoStop}%
\bibitem [{\citenamefont {Franklin}\ \emph {et~al.}(1998)\citenamefont {Franklin}, \citenamefont {Powell}, \citenamefont {Workman} \emph {et~al.}}]{franklin1998digital}%
  \BibitemOpen
  \bibfield  {author} {\bibinfo {author} {\bibfnamefont {G.~F.}\ \bibnamefont {Franklin}}, \bibinfo {author} {\bibfnamefont {J.~D.}\ \bibnamefont {Powell}}, \bibinfo {author} {\bibfnamefont {M.~L.}\ \bibnamefont {Workman}}, \emph {et~al.},\ }\href@noop {} {\emph {\bibinfo {title} {Digital control of dynamic systems}}},\ Vol.~\bibinfo {volume} {3}\ (\bibinfo  {publisher} {Addison-wesley Menlo Park},\ \bibinfo {year} {1998})\BibitemShut {NoStop}%
\bibitem [{\citenamefont {Lewis}\ \emph {et~al.}(2017)\citenamefont {Lewis}, \citenamefont {Xie},\ and\ \citenamefont {Popa}}]{lewis2017optimal}%
  \BibitemOpen
  \bibfield  {author} {\bibinfo {author} {\bibfnamefont {F.~L.}\ \bibnamefont {Lewis}}, \bibinfo {author} {\bibfnamefont {L.}~\bibnamefont {Xie}},\ and\ \bibinfo {author} {\bibfnamefont {D.}~\bibnamefont {Popa}},\ }\href@noop {} {\emph {\bibinfo {title} {Optimal and robust estimation: with an introduction to stochastic control theory}}}\ (\bibinfo  {publisher} {CRC press},\ \bibinfo {year} {2017})\BibitemShut {NoStop}%
\bibitem [{\citenamefont {Whittle}\ \emph {et~al.}(2021)\citenamefont {Whittle}, \citenamefont {Hall}, \citenamefont {Dwyer}, \citenamefont {Mavalvala}, \citenamefont {Sudhir} \emph {et~al.}}]{whittle_approaching_2021}%
  \BibitemOpen
  \bibfield  {author} {\bibinfo {author} {\bibfnamefont {C.}~\bibnamefont {Whittle}}, \bibinfo {author} {\bibfnamefont {E.~D.}\ \bibnamefont {Hall}}, \bibinfo {author} {\bibfnamefont {S.}~\bibnamefont {Dwyer}}, \bibinfo {author} {\bibfnamefont {N.}~\bibnamefont {Mavalvala}}, \bibinfo {author} {\bibfnamefont {V.}~\bibnamefont {Sudhir}}, \emph {et~al.},\ }\bibfield  {title} {\bibinfo {title} {Approaching the motional ground state of a 10-kg object},\ }\href {https://doi.org/10.1126/science.abh2634} {\bibfield  {journal} {\bibinfo  {journal} {Science}\ }\textbf {\bibinfo {volume} {372}},\ \bibinfo {pages} {1333} (\bibinfo {year} {2021})}\BibitemShut {NoStop}%
\bibitem [{\citenamefont {Bereyhi}\ \emph {et~al.}(2022)\citenamefont {Bereyhi}, \citenamefont {Arabmoheghi}, \citenamefont {Beccari}, \citenamefont {Fedorov}, \citenamefont {Huang}, \citenamefont {Kippenberg},\ and\ \citenamefont {Engelsen}}]{bereyhi2022perimeter}%
  \BibitemOpen
  \bibfield  {author} {\bibinfo {author} {\bibfnamefont {M.~J.}\ \bibnamefont {Bereyhi}}, \bibinfo {author} {\bibfnamefont {A.}~\bibnamefont {Arabmoheghi}}, \bibinfo {author} {\bibfnamefont {A.}~\bibnamefont {Beccari}}, \bibinfo {author} {\bibfnamefont {S.~A.}\ \bibnamefont {Fedorov}}, \bibinfo {author} {\bibfnamefont {G.}~\bibnamefont {Huang}}, \bibinfo {author} {\bibfnamefont {T.~J.}\ \bibnamefont {Kippenberg}},\ and\ \bibinfo {author} {\bibfnamefont {N.~J.}\ \bibnamefont {Engelsen}},\ }\bibfield  {title} {\bibinfo {title} {Perimeter modes of nanomechanical resonators exhibit quality factors exceeding 10 9 at room temperature},\ }\href@noop {} {\bibfield  {journal} {\bibinfo  {journal} {Physical Review X}\ }\textbf {\bibinfo {volume} {12}},\ \bibinfo {pages} {021036} (\bibinfo {year} {2022})}\BibitemShut {NoStop}%
\bibitem [{\citenamefont {Norte}\ \emph {et~al.}(2016)\citenamefont {Norte}, \citenamefont {Moura},\ and\ \citenamefont {Gr{\"o}blacher}}]{norte2016mechanical}%
  \BibitemOpen
  \bibfield  {author} {\bibinfo {author} {\bibfnamefont {R.~A.}\ \bibnamefont {Norte}}, \bibinfo {author} {\bibfnamefont {J.~P.}\ \bibnamefont {Moura}},\ and\ \bibinfo {author} {\bibfnamefont {S.}~\bibnamefont {Gr{\"o}blacher}},\ }\bibfield  {title} {\bibinfo {title} {Mechanical resonators for quantum optomechanics experiments at room temperature},\ }\href@noop {} {\bibfield  {journal} {\bibinfo  {journal} {Physical Review Letters}\ }\textbf {\bibinfo {volume} {116}},\ \bibinfo {pages} {147202} (\bibinfo {year} {2016})}\BibitemShut {NoStop}%
\bibitem [{\citenamefont {Bachtold}\ \emph {et~al.}(2022)\citenamefont {Bachtold}, \citenamefont {Moser},\ and\ \citenamefont {Dykman}}]{bachtold2022mesoscopic}%
  \BibitemOpen
  \bibfield  {author} {\bibinfo {author} {\bibfnamefont {A.}~\bibnamefont {Bachtold}}, \bibinfo {author} {\bibfnamefont {J.}~\bibnamefont {Moser}},\ and\ \bibinfo {author} {\bibfnamefont {M.}~\bibnamefont {Dykman}},\ }\bibfield  {title} {\bibinfo {title} {Mesoscopic physics of nanomechanical systems},\ }\href@noop {} {\bibfield  {journal} {\bibinfo  {journal} {Reviews of Modern Physics}\ }\textbf {\bibinfo {volume} {94}},\ \bibinfo {pages} {045005} (\bibinfo {year} {2022})}\BibitemShut {NoStop}%
\bibitem [{\citenamefont {Wiener}(1930)}]{wiener1930generalized}%
  \BibitemOpen
  \bibfield  {author} {\bibinfo {author} {\bibfnamefont {N.}~\bibnamefont {Wiener}},\ }\bibfield  {title} {\bibinfo {title} {Generalized harmonic analysis},\ }\href@noop {} {\bibfield  {journal} {\bibinfo  {journal} {Acta mathematica}\ }\textbf {\bibinfo {volume} {55}},\ \bibinfo {pages} {117} (\bibinfo {year} {1930})}\BibitemShut {NoStop}%
\bibitem [{\citenamefont {Zhang}\ and\ \citenamefont {St-Gelais}(2023)}]{zhang2023demonstration}%
  \BibitemOpen
  \bibfield  {author} {\bibinfo {author} {\bibfnamefont {C.}~\bibnamefont {Zhang}}\ and\ \bibinfo {author} {\bibfnamefont {R.}~\bibnamefont {St-Gelais}},\ }\bibfield  {title} {\bibinfo {title} {Demonstration of frequency stability limited by thermal fluctuation noise in silicon nitride nanomechanical resonators},\ }\href@noop {} {\bibfield  {journal} {\bibinfo  {journal} {Applied Physics Letters}\ }\textbf {\bibinfo {volume} {122}} (\bibinfo {year} {2023})}\BibitemShut {NoStop}%
\bibitem [{\citenamefont {Doyle}\ \emph {et~al.}(2013)\citenamefont {Doyle}, \citenamefont {Francis},\ and\ \citenamefont {Tannenbaum}}]{doyle2013feedback}%
  \BibitemOpen
  \bibfield  {author} {\bibinfo {author} {\bibfnamefont {J.~C.}\ \bibnamefont {Doyle}}, \bibinfo {author} {\bibfnamefont {B.~A.}\ \bibnamefont {Francis}},\ and\ \bibinfo {author} {\bibfnamefont {A.~R.}\ \bibnamefont {Tannenbaum}},\ }\href@noop {} {\emph {\bibinfo {title} {Feedback control theory}}}\ (\bibinfo  {publisher} {Courier Corporation},\ \bibinfo {year} {2013})\BibitemShut {NoStop}%
\end{thebibliography}%

\end{document}